\documentclass[%
 reprint,
 amsmath,amssymb,
 aps,
]{revtex4-1}

\usepackage{graphicx}
\usepackage{dcolumn}
\usepackage{bm}
\usepackage{makecell}
\usepackage{array,multirow}

\begin{document}

\preprint{APS/123-QED}

\title{Fluid flow through packings of elastic shells}

\author{Pawel Gniewek}
 \email{pawel.gniewek@berkeley.edu}
 \affiliation{%
Biophysics Graduate Group, University of California, Berkeley, USA
}

\author{Oskar Hallatschek}%
 \email{ohallats@berkeley.edu}
\affiliation{%
Departments of Physics and Integrative Biology, University of California, Berkeley, USA
}

\hyphenation{Review}
\hyphenation{Physical}

\date{\today}

\begin{abstract}
Fluid transport in porous materials is commonly studied in geological samples (soil, sediments etc.) or idealized systems, but the fluid flow through compacted granular materials, consisting of substantially strained granules, remains relatively unexplored. As a step towards filling this gap, we study a model of liquid transport in packings of deformable elastic shells using Finite Element and Lattice-Boltzmann methods. We find that the fluid flow abruptly vanishes as the porosity of the material falls below a critical value, and the flow obstruction exhibits features of a percolation transition.
We further show that the fluid flow can be captured by a simplified permeability model in which the complex porous material is replaced by a collection of disordered capillaries, which are distributed and shaped by the percolation transition. To that end, we numerically explore the divergence of hydraulic tortuosity $\rm\tau_H$ and the decrease of a hydraulic radius $\rm R_h$ as the percolation threshold is approached. We interpret our results in terms of scaling predictions derived from the percolation theory applied to random packings of spheres.

\end{abstract}

\pacs{Valid PACS appear here}
\maketitle

\section{INTRODUCTION}
The physics of fluid flow through disordered porous media is of fundamental importance to a wide range of engineering and scientific fields including enhanced oil recovery, carbon capture and storage, contamination migration in ground-water, water transport, and nutrient transport in tissues and microbial colonies \cite{sahimi1993flow, bear2010,sahimi2011,delarue2016self,hornung2018quantitative}.
This has led to a substantial effort in looking for relationships between the effective physical transport properties and the structural properties of porous materials. In spite of the extensive work that has been done, a full description of liquid transport in a broad range of material parameters is elusive \cite{hunt2017flow}. Experimental studies, especially in 3D systems, are limited because imaging material samples and resolving fluid flow stream lines are challenging tasks \cite{gaillard2007imaging, gostovic2007three, scholtz2012permeability, tahmasebi2017image}.
Numerical studies are most often tackled in 2D due to the high computational burden \cite{cancelliere1990permeability,koponen1996tortuous,koponen1997permeability,matyka2008tortuosity,koza2009finite,duda2011hydraulic}. Even though a broad range of material porosities in 2D systems has been covered, a drawback of these studies is that, for disordered materials, the percolation transition coincides with the rigidity transition \cite{zallen84}.
For 3D systems, simulations are commonly performed for an idealized model of randomly distributed inter-penetrating objects like cubes or spheres \cite{maier1998simulation,stewart2006study,matyka2012calculate,bakhshian2016computer}. These systems are good prototypes to study critical phenomena, but liquid transport in complex geometries depends on boundary condition details; thus, the relevance of these models for actual materials is not clear \cite{lehmann2008impact}. 
There is also work done on fluid transport in geometries obtained from the microtomography of collected materials. However, these studies are performed usually for a small number of samples and at relatively high porosity \cite{chen2008pore,chen2009temporal}.

In recent years, the interest in granular systems made of deformable and strongly compacted elastic shells and membranes increased \cite{jose2015jammed,jose2016random,boromand2018jamming}. This class of models is of interest not only in physics and engineering, but also increasingly in biological research of small cell clusters \cite{sandersius2008modeling,milde2014sem}, epithelial cells \cite{jamali2010subcellular}, and jammed microbial packings in confined spaces \cite{delarue2016self,gniewekphdthesis}. Henceforth, in this work, we focus on a 3D model of granular materials where particles are represented as elastic spherical shells, with the volume of these shells kept constant (motivated by experimental work on confined microbial populations \cite{delarue2016self,gniewekphdthesis}). For such a model of the granular system, we numerically study a single-phase viscous flow in Darcy's regime, {\it i.e.} laminar flow with a linear relation between volumetric flow and pressure gradient.
We consider packings in a broad range of porosities, from the point the packings start to be mechanically stable (jamming transition \cite{o2003jamming}) down to the porosities where the liquid transport ceases to exist (percolation transition \cite{zallen84}). 
We mainly focus on a model by Kozeny and Carman \cite{kozeny1927uber,carman1937fluid} \textemdash~the classical permeability-porosity framework. First, we briefly introduce the Kozeny-Carman model. Then, we present how the key features of the Kozeny-Carman model can be physically grounded in a percolation theory. Finally, we present numerical evidence on how different structural features of granular porous material contribute to the fluid transport in granular porous media. 

\subsubsection*{Kozeny-Carman Model}
Permeability $\kappa$ measures the ability of fluid to flow through porous media and it is part of the proportionality constant in Darcy's law, the relation between the fluid volumetric flux U (discharge per unit area) and a pressure gradient:
\begin{equation}\label{eq:darcy}
\rm U = -\frac{\kappa}{\eta} \nabla P(\mathbf{r})
\end{equation}
where U is given in units of length/time, $\eta$ is the dynamic viscosity of the fluid, and $\rm P(\mathbf{r})$ is the pressure at the location $\mathbf{r}$. This phenomenological relation is valid at low Reynolds numbers when the flow is laminar. For small pressure gradients, we can further assume $\rm\nabla P = \Delta P / \rm L$, where $\rm L$ is the linear size of the system.
\\

\noindent
For low Reynolds number flow in a straight and cylindrical capillary channel, the volumetric flux is given by the Poiseuille equation:
\begin{eqnarray}
\rm U = -B \frac{R^2}{\eta} \frac{ \Delta P }{\rm L}
\end{eqnarray}
where $\rm R$ is the radius of a capillary, $\rm B$ is a numerical factor, and $\rm L$ is the length of the capillary.
If a capillary occupies only a fraction of the material, the liquid discharge per area unit is correspondingly lower. Assuming that the capillaries are homogeneously distributed in the material, the scaling factor is the amount of the void space in the material, called a porosity $\phi$:
\begin{eqnarray}\label{Ueq}
\rm U = -\phi B \frac{R^2}{\eta} \frac{ \Delta P }{\rm L}
\end{eqnarray}

For capillaries that are not straight, Kozeny pointed out that due to the tortous character of the flow, the length of the equivalent channels should be $\rm\left<\lambda\right> \equiv \tau_H \cdot L$, where $\rm\tau_H$ is called hydraulic tortuosity, and the fluid discharge needs to be scaled down by it \cite{kozeny1927uber}. Carman further reasoned that it takes $\rm\tau_H$ times more time to discharge the same amount of fluid through porous media than it takes for straight capillaries (in a macroscopic direction of the flow). Thus, the discharge rate should additionally be $\rm\tau_H$-times smaller \cite{carman1937fluid}. Capillaries are not limited to just the circular cross-sections. For the general shape of the capillary, the radius R is commonly replaced by a hydraulic radius $\rm R_h$ \cite{costa2006permeability} (defined as the ratio of the cross-sectional area normal to flow to the wetted perimeter of the flow channels), but sometimes other parameters are used, for example, the critical pore radius \cite{nishiyama2017permeability}.

Thus from Equation \ref{Ueq}, the final relation for the capillary flow in a porous material is \cite{carman1937fluid}:
\begin{equation} \label{eq:flow}
\rm U = -B \frac{\phi R_h^2}{\tau_H\cdot\eta} \cdot \frac{ \Delta P }{\rm \tau_H \cdot L} = -B\frac{\phi R_h^2}{\tau_H^2} \frac{1}{\eta} \nabla P
\end{equation}

\noindent
Comparing Equation \ref{eq:darcy} with Equation \ref{eq:flow}, a general formula for permeability reads
\begin{equation}\label{eq:kappa}
\rm\kappa = B\frac{\phi R_h^2}{\tau_H^2}
\end{equation}
and is called the Kozeny-Carman equation. Despite being semi-empirical, Equation \ref{eq:flow} is commonly used as a simple model for the permeability in porous materials.

\section{METHODS}
\subsection{Packings of Deformable Shells}\label{packingsgeneration}

\subsubsection{Generation of Compressed Packings}
The initial packings of the shells are generated using a standard jamming, with a periodic boundary conditions algorithm \cite{o2003jamming}. Starting from these jammed packings, more compacted packings are generated by changing a linear dimension of the simulation box. The changes of the box size are minute, and less than 0.4\% of the size of an elastic shell. After every box size change, the mechanical stresses are relaxed using the FIRE algorithm \cite{bitzek2006structural}, see Section \ref{appx:jammig} fore more details.

\subsubsection{Shells Mechanics}
Every shell is modeled as a membrane using about 5000 triangular finite elements per shell. The ratio of a shell thickness $\rm t$ to the initial diameter $\rm D_0$ is $\rm t/\rm D_0 = 0.02$, so bending effects can be neglected and the shell material is modeled as an isotropic St. Venant-Kirchhoff membrane \cite{delingette2008triangular,delingette2008biquadratic}. All of the shells are slightly pressurized at the beginning of the simulation, with initial pressure $\rm P_0$, and filled with an incompressible liquid. The ratio between $\rm P_0$ and the Young's modulus of the shell $\rm E$ is equal to $\rm P_0/\rm E = 0.0025$.
The force due to the shell volume-dependent pressure $\rm P(V_{\rm shell})$ on a vertex 
$i$ is calculated as:
$\rm F(\mathbf{r}_i)=\nabla_{\mathbf{r}_i}\big(\rm P(V_{\rm shell})\cdot V_{\rm shell}\big)$
where $\rm V_{\rm shell}(\mathbf{r}_1,...,\mathbf{r}_{N_{\rm vert})}$ is a function of the $N_{\rm vert}$
vertices in the mesh and the volume change for the vertex $i$ is calculated using the tetrahedral volume defined by the vertex $i$, its neighboring vertices in the mesh, and the center of the mass \cite{delarue2016self,gniewekphdthesis}.
Once the mechanical forces are equilibrated, the constant shell volume constraint is enforced by varying the shells' internal pressures. If the volume of a shell is not equal to the preassigned value $\rm V_0$, the pressure is adjusted to the new value $\rm P_{\rm new} = \rm P_{\rm old} \left[1 + \left(\rm V_0-\rm V\right)/\rm V\right]$. This inevitably drags the system out of mechanical equilibrium and the system needs to be equilibrated again. The protocol continues until the volumes of the shells reach their preassigned volumes within 0.1\% of accuracy \cite{gniewekphdthesis}.

\subsection{Identification of a Percolating Cluster}\label{percoprotocol}
\begin{figure}
\includegraphics[width=0.9\columnwidth,keepaspectratio]{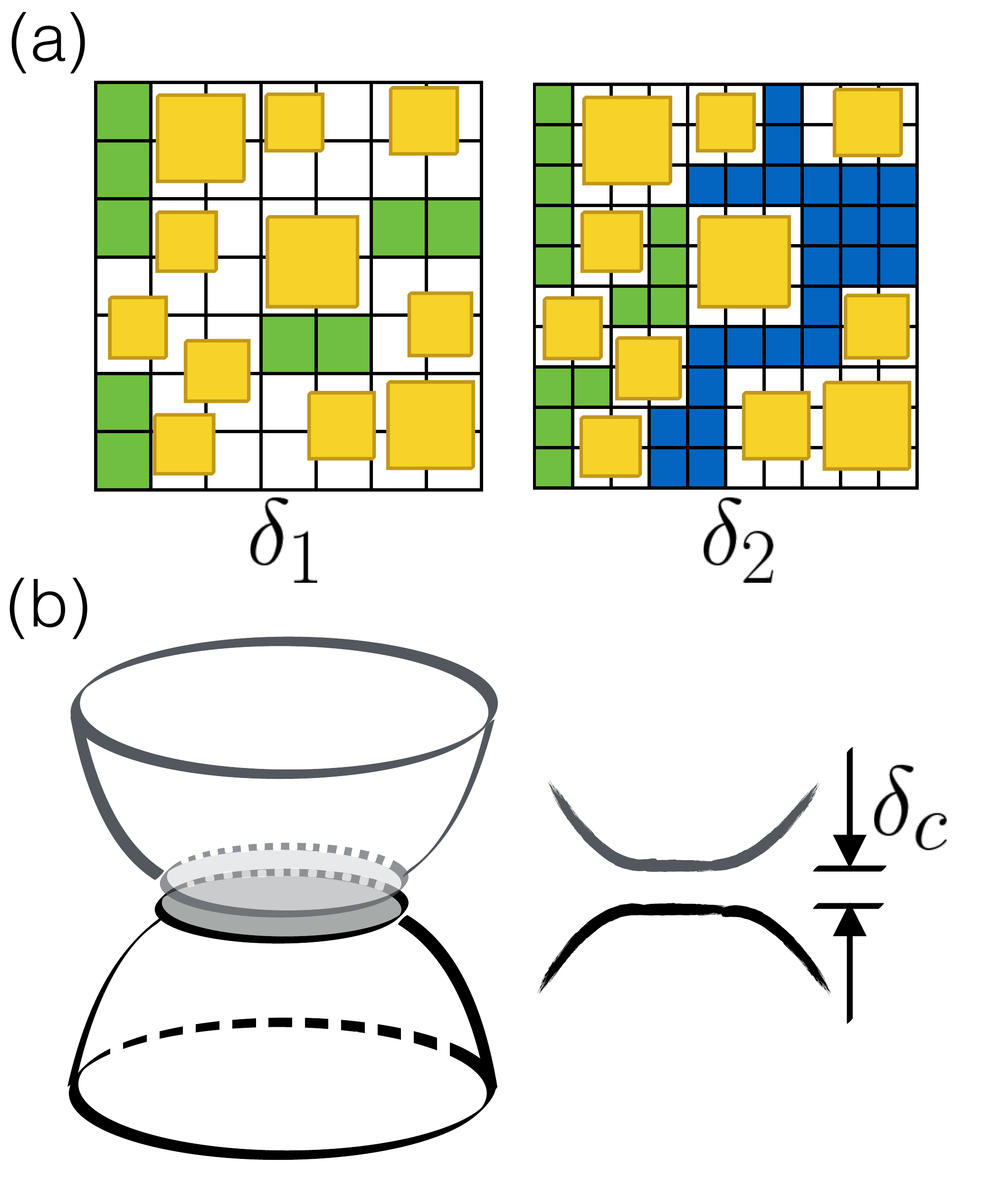}
\caption{\textbf{Identification of a percolating cluster}: \textbf{(a)} A schematic of a 2D system projected onto a lattice with two different lattice sizes. Every lattice site that contains any part of a particle (yellow shapes) is considered to be occupied and impermeable to liquid (white squares). For the lattice resolution $\delta_1$ and a given configuration, there is no percolating cluster capable of carrying liquid through the packing, but there are some unoccupied lattice sites (green squares). For the lattice size $\delta_2$, there is a percolating cluster (blue squares). There are also some unoccupied lattice sites (green squares) that do not belong to the percolating cluster. \textbf{(b)} A sketch of two elastic shells in contact. Shells are represented by mid-surfaces, so despite the fact that they are in contact, there is a finite gap between them ($\delta_c\approx 0.025$).}
\label{fig:lattice}
\end{figure}

To identify clusters that percolate the void space between the shells, we project a packing of shells onto a 3D lattice with a lattice constant $\delta$, see Fig.~\ref{fig:lattice}a. Every lattice site that contains a shell's vertex is considered impermeable to the liquid, Fig.~\ref{fig:lattice}a.
The shells are represented as finite elements. Thus, for a small enough lattice constants $\delta$, the membrane is permeable to the liquid, {\it i.e.} the liquid can enter the interior of the shell. This problem can be overcome by identifying impermeable lattice sites using triangles defined by vertices rather than by vertices alone. 
However, the mid-surface plane is used to represent the three-dimensional shells in two-dimensional form, so even though two shells are in contact, there is a finite gap between their mid-surfaces, Fig.~\ref{fig:lattice}b. Thus, below a certain lattice size $\delta_c \approx 0.025$, the packings can be permeable due to this finite gap, and percolating clusters identified for $\delta < \delta_c$ are dubious.

Finally, we look for a percolating cluster using the connected-component labeling algorithm (implemented in the \texttt{scipy.ndimage} Python library). The cluster is said to percolate the system if it contains lattice sites on the two opposite sides of the simulation box.
One of the characteristic length-scales in the system is the initial diameter of a shell, $D_0$. We choose to express the lattice sizes, $\delta$, in units of $D_0$. In principle, we would like to generate a lattice with $\delta \rightarrow 0$ as we want to estimate a fluid flow in the continuum limit. However, due to the aforementioned limitations, the resolution of the lattices in our study is finite and varies from a coarse one to a fine one, and it is in the range [0.03, 0.07].
Finally, percolation clusters identified in this way are used for hydraulic radius and Lattice-Boltzmann calculations. 

\subsection{Lattice Boltzmann Simulations} \label{lbsims}
Velocity fields of the fluid flow through the packings of the shells are solved with the Lattice-Boltzmann (LB) method \cite{succi2001lattice} using the D3Q19 lattice. This method was proven to be successful in studies of liquid flow in porous materials \cite{cancelliere1990permeability,koponen1996tortuous, koponen1997permeability, maier1998simulation,arns2005cross, stewart2006study, pan2006evaluation, chen2008pore, chen2009temporal, matyka2008tortuosity, duda2011hydraulic, matyka2012calculate, bakhshian2016computer, jin2016statistics, singh2017impact, amarsid2017viscoinertial}. We use this method to obtain a solution to the Navier-Stokes equation for the flow in low Reynolds numbers limit. The LB method is using a velocity distribution function rather than velocity and pressure fields and is numerically more stable than the Finite Element Method at the irregular boundaries that are inevitable in porous materials \cite{succi2001lattice}. 
To ensure better numerical stability for the complex geometry of the pores, we use multiple relaxation times (MRT) to solve linearized Boltzmann equation with LB method \cite{premnath2007three}.

Permeability of the packing and the flow field are resolved by setting a pressure difference $\rm\Delta P$ between two opposite sides of the simulation box, sufficiently small to keep the flow in the incompressible and laminar regimes (Stokes flow). 
Every simulation is performed for periodic boundary condition (PBC) in directions perpendicular to the pressure gradient. In the direction of the pressure gradient, the system is open and the boundary conditions are set by pressure difference \cite{chen2008pore,chen2009temporal}.
No-slip boundary condition is applied to the solid material boundaries.
It has been found \cite{succi2001lattice,matyka2008tortuosity} that when the channels carrying liquid become very narrow (of the order of one lattice site) LB simulations become unstable and the evaluation of the stream lines become inaccurate. To deal with this problem we use an approach proposed by Matyka {\it et al.} \cite{matyka2008tortuosity}, where every lattice site on which flow equations are solved is further refined into $\rm M^3$ smaller cubic elements (refinement level: M). Strictly speaking, $\tilde{\delta}=\delta/\rm M$ is a lattice size of the fluid phase, and throughout this paper we use $\rm M=3$ (unless stated otherwise). Due to computational limitations, LB calculations are performed for the lattice constant $\delta = 0.04$ (unless stated otherwise). 

The flow fields obtained from LB simulations for each lattice site, $u(\mathbf{r})$, are further used to calculate the permeability and the tortuosity. 
Permeability is calculated as $\kappa = \eta \cdot \left<u(\mathbf{r})\right> / \nabla P$, and tortuosity as $\tau_H = \frac{\left<u(\mathbf{r})\right>}{\left<u(\mathbf{r})_{x}\right>}$, see Section \ref{appx:tau_calc} for the formal derivation. Permeability is given in lattice units (for conversion to physical units follow Latt \cite{latt2008choice}).
All the LB simulations are performed with PALABOS (\url{http://www.palabos.org}).

The simulated model of the porous material accounts for deformability and the mechanics of the shell membrane using Finite Element method. The mechanics of the shells are resolved with $\sim 3.75\cdot 10^5$ degrees of freedom, and some of the LB simulations required up to $\sim 10^7$ lattice points to resolve the fluid velocity field. In turn, the resolution of the calculation imposes restrictions on the largest system size that we are able to study. Finite size effects for the studied systems may result in small anisotropies in the permeability tensor \cite{koza2009finite}, but recent studies show that transport in complex porous geometries can be reasonably well captured if the size of the system is roughly $\gtrsim 10$ times larger than the pore size \cite{arns2005cross,chen2008pore, jager2017mechanism, jager2017channelization}.

\section{RESULTS}

\subsection{Percolation Transition} \label{percothreshold}
\begin{figure*}
\includegraphics[width=\textwidth,keepaspectratio]
{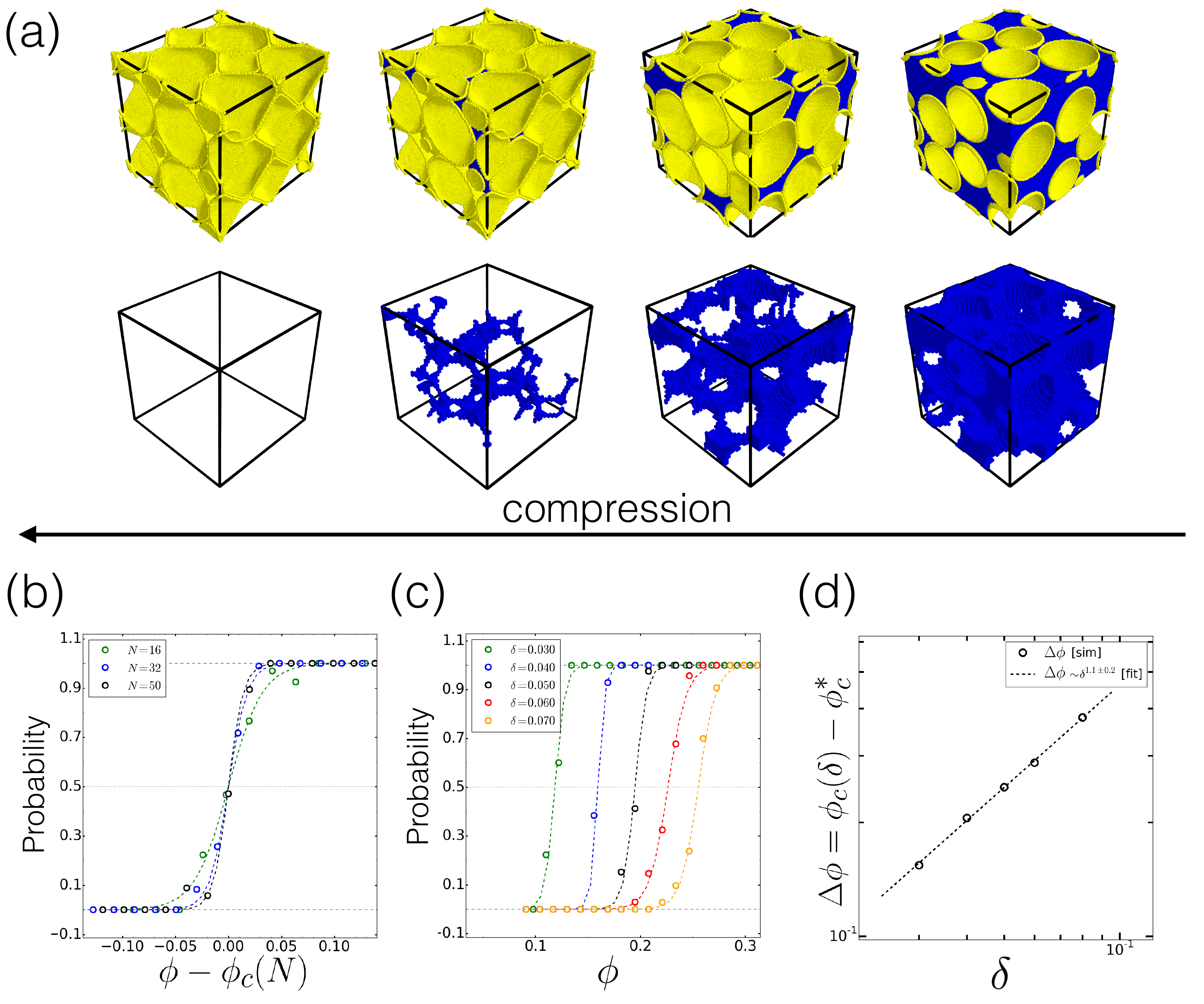}
\caption{\textbf{Percolation transition}:  \textbf{(a)} Percolating clusters (in blue) and deformable shells (in yellow) in the packing with periodic boundary conditions as the compaction of the system progresses. For clarity, the smallest system is presented (N=16) with a lattice resolution $\delta=0.04$. As the system is more and more compact, the percolating cluster gets smaller and more tortuous, and eventually disappears at the critical porosity. \textbf{(b)} Percolation probability for three system sizes: N=16, 32, and 50. Dashed lines are sigmoid fits to the numerical data, and binned averages are given by open dots. The lattice-size dependent percolation threshold $\phi_c(N)$ has been estimated as the porosity value for which percolation probability is equal to 0.5. The plots represent data for the lattice $\delta=0.04$. \textbf{(c)} Percolation probabilities for the system size N=50 and varying lattice sizes: $\delta = \{0.03, 0.04, 0.05, 0.06, 0.07\}$. As the resolution of the lattice increases, the percolation threshold shifts toward lower porosity values. \textbf{(d)} We extrapolated the percolation threshold in the continuum limit. Dashed line is a power-law fit, where $\phi^*_c$ and the exponent for $\delta$ are two fitting parameters. The fitted percolation threshold in continuum limit is $\phi_c^* = 0.035\pm 0.014$, and the exponent is equal to $1.1\pm 0.2$. The relation $\Delta \phi \sim \delta^{1.1}$ agrees well with the work of Koza {\it et al}. \cite{koza2014percolation}. Details of a fitting procedure can be found in Section \ref{fitting_protocols}.}
\label{fig:perctrans}
\end{figure*}

In idealized systems, such as random packings of overlapping cubes or spheres (and their complementaries, where the solid material is drilled in random locations, rather than deposited \cite{halperin1985differences,halperin1987transport}), the void space between them undergo a percolation transition \cite{kerstein1983equivalence,elam1984critical,van1996network,koza2014percolation,priour2014percolation}. Since, in the vicinity of the percolation threshold, a minute deposition of solid material can disconnect the percolating cluster and prevent further liquid transport, the abruptness of this transition is well understood. The model studied in this work differs from the aforementioned ones in that the narrow necks in the percolating cluster decay continuously upon the compaction of the material. It is however akin to the cherry-pit model \cite{torquato2002random}, where the sizes of impermeable obstacles are continuously increased \cite{schnyder2015rounding} and long-time transport properties vanish at the percolation threshold due to an underlying continuum percolation transition of the liquid accessible space \cite{schnyder2015rounding}.
To study this aspect in our model, following the protocol described in Section \ref{percoprotocol}, percolating clusters are identified for three system sizes (N=16, 32, and 50 elastic shells) and various lattice resolutions, Fig.~\ref{fig:perctrans}a. The results for $\delta = 0.04$ are shown in Fig.~\ref{fig:perctrans}b. As the system gets larger, the transition becomes steeper, as expected in a first-order transition case \cite{elam1984critical,priour2014percolation,soltani2018scaling}. The steepness of this transition depends on the system size $L$, and scales as $\sim L^{1/\nu}$, where $\nu$ is a critical exponent of the correlation length. In a continuum percolation model, this exponent is approximately equal $\nu \approx 0.88$ \cite{SA94}.

Fig.~\ref{fig:perctrans}c shows that a sharp drop in fluid transport capabilities occurs for different lattice resolutions and that the percolation threshold shifts towards lower porosity values as $\delta$ decreases \textemdash~an effect anticipated from the studies on idealized lattice models \cite{koza2014percolation}.
The finite representation of the elastic shells in the studied model does not allow for calculations in the continuum limit. It is nevertheless possible to extrapolate a percolation threshold in the continuum limit $\delta \rightarrow 0$. In Fig.~\ref{fig:perctrans}d, we estimated that for N=50, the percolation threshold in the continuum limit is $\phi_c^*(\rm N=50)=0.035\pm 0.014$, which is consistent with the values obtained for other granular porous materials \cite{van1996network,priour2014percolation,liu211application,koza2014percolation,berg2017formation}.
\noindent

For each system size, the percolation threshold $\phi_c$ for a finite $\delta$ and $N$ is expected to be related to the threshold in the continuum limit $\phi_c^*$ as a power-law $\phi_c(N)-\phi_c^*(N)\equiv\Delta\phi(N) \sim \delta^{\beta}$ \cite{koza2014percolation}. In Fig.~\ref{fig:perctrans}d, we estimate the lattice-size scaling exponent to be $\beta=1.1$ for N=50, and similar values of $\beta$ are found for $N=16, 32$; {\it cf.} Table~\ref{tab:fits}. The value of the exponent $\beta$ is in good agreement with the prediction made by Koza {\it et al.} \cite{koza2014percolation}, where the exponent is estimated to be $\beta\approx 1$ \textemdash~yielding an approximate relation for the lattice-size dependent percolation threshold that obeys: $\phi_c(N) -\phi_c^*(N) \sim \delta$. Additionally, these fits in the continuum limit are subject to a finite system size correction that overestimates (in a first order) the thermodynamic limit by $\phi_c^*(N) = \phi_c^*(N\rightarrow\infty) + C_{I}L^{-1/\nu}$, where $C_I\sim \mathcal{O}(1)$, $L\sim N^{1/d}$, and $d=3$ \cite{lorenz1993universality,van1996network,rintoul1997precise,liu211application}. An accurate extrapolation to the thermodynamic limit requires data for systems spanning many orders of magnitude, but in Section \ref{taudivergence} and Section \ref{karmancozenymodel} we show that transport properties discussed in this work do not depend on the exact value of $\phi_c^*(N\rightarrow\infty)$, but rather on a reduced porosity $\delta\phi(\delta,N)\equiv\phi - \phi_c(\delta,N)$ \textemdash~a value that can be well estimated for a given lattice size $\delta$, and system size $N$ \cite{martys1994universal}.
The model presented in this contribution can be further detailed, but the numerical results clearly point to common characteristics between the model studied in this work and previously studied percolation models \cite{martys1994universal,van1996network,rintoul1997precise,priour2014percolation,koza2014percolation,schnyder2015rounding}. Thus, we use the formalism of percolation theory in the analysis of fluid flow obstruction in the vicinity of the critical porosity value $\phi_c$, which in this study is $\phi_c\approx 0.15$ (unless stated otherwise). 

\subsection{Decrease of Hydraulic Radius $R_h$ with the Porosity} \label{rhresults}
The hydraulic radius is defined as a ratio of a cross-section of a liquid carrying channel to its wetted perimeter, see Section \ref{appx:hydraulicradius} for more details. Only in relatively simple cases, such as a laminar flow inside a pipe, can the hydraulic radius be directly related to the geometry of the system. In practice, finding this value is problematic because it is difficult to accurately predict a channel's shape along the flow stream lines. The situation gets even more complicated in complex geometries where percolating channels can merge or branch out. Thus, the hydraulic radius is commonly approximated by the ratio of the volume to the wetted area of a cluster carrying the liquid \cite{clennell1997tortuosity}.

\begin{figure}
\includegraphics[width=\columnwidth,keepaspectratio]{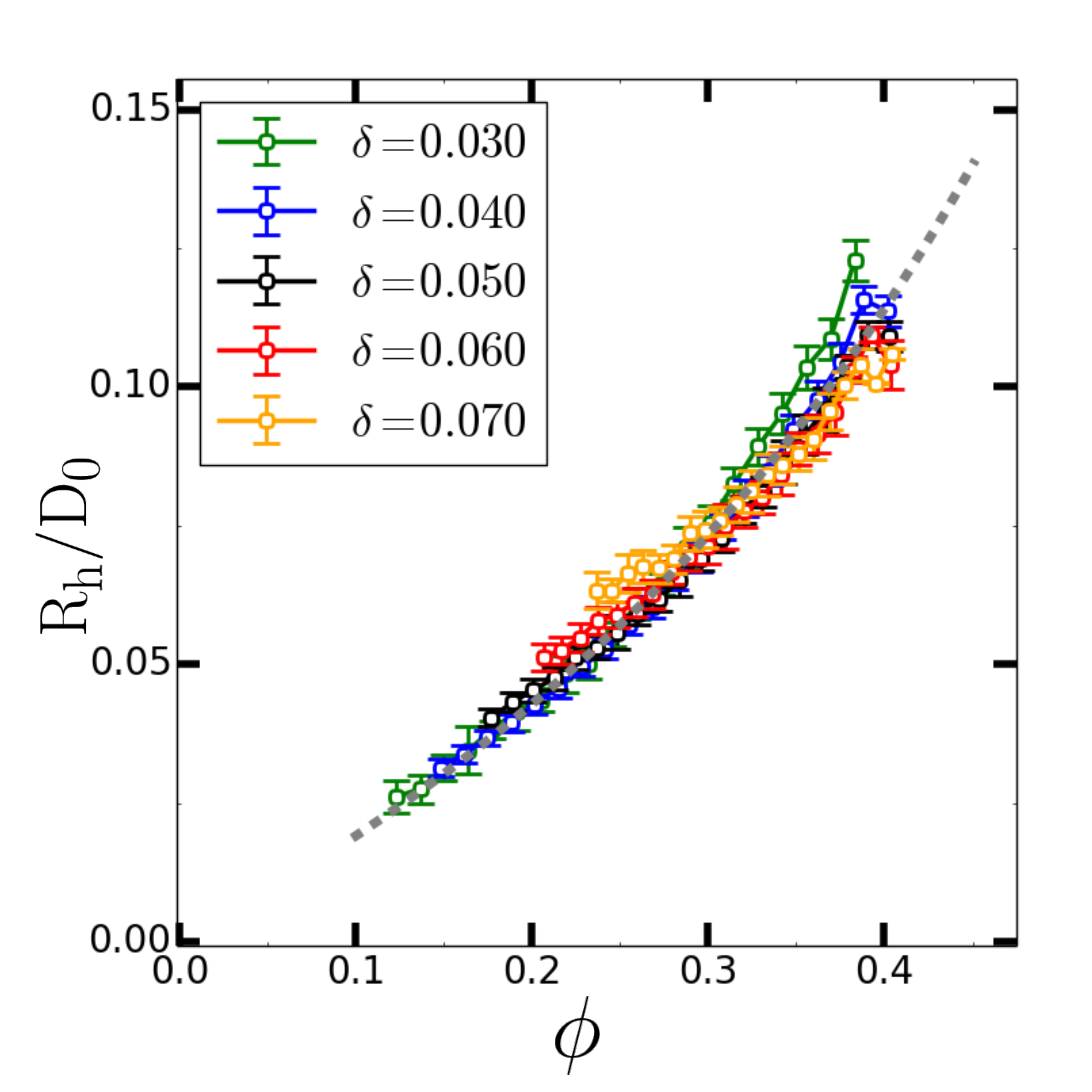}
\caption{\textbf{Hydraulic radius $R_h$}: Hydraulic radius as a function of porosity $\phi$ for the system size N=50. For each of the identified percolating cluster, the hydraulic radius was calculated as a ratio of the cluster volume to the surface area, and then normalized by the initial size of an elastic shell $D_0$. Error bars give one standard deviation. Dashed-lines are the fits to $R_h\propto \phi/(1-\phi)$ for $\delta=0.04$. At the jamming point ($\phi\approx 0.4$) for monodisperse packings of shells with the size $D_0$, the numerical data is in agreement with the expected theoretical value $R_h/D_0\approx 0.14$. See Section \ref{appx:hydraulicradius} for more details.}
\label{fig:Rhinv}
\end{figure}

Using the percolating clusters identified for the packings of elastic shells, we estimated the hydraulic radii for different lattice resolutions as a ratio of the number of lattice sites belonging to the cluster divided by the number of surface sites \cite{arns2001euler,arns2005cross}. Using a geometric argument adapted from references \cite{ng1986model,macdonald1991generalized}, the hydraulic radius is predicted to vanish linearly at the limit of zero porosity, see Section \ref{appx:hydraulicradius}. Results corroborating this prediction can be found in Fig.~\ref{fig:Rhinv}. The results indicate that the hydraulic radius decays like:
\begin{equation}\label{rinvresult}
\rm R_h \propto \phi/(1-\phi)
\end{equation}
as the porosity goes to 0. If the hydraulic radius was reaching 0 at the percolation threshold $\phi_c$, this would indicate that as the porosity approaches the percolation threshold $\phi\rightarrow\phi_c$, most of the fluid flow occurs in the layer in the vicinity of the percolating cluster's bounding surface, where the effects of viscosity are significant. Thus, the liquid transport could be controlled by a no-slip boundary condition on the cluster's surface and not necessarily the complex geometry of the cluster. However, the hydraulic radius vanishes independently of lattice size, and its value at the percolation threshold is finite, as one would expect from a percolation theory \cite{leath1978scaling}.

\subsection{Tortuosity Divergence at the Percolation Threshold}\label{taudivergence}
Tortuosity underpins the relationship between a transport process and the underlying geometry and topology of the pores \cite{yu2002fractal}.
Recently it has been shown that the tortuosity depends on material structural properties, and may vary significantly close to the percolation threshold \cite{costa2006permeability, matyka2008tortuosity,duda2011hydraulic,ahmadi2011analytical,ahmadi2012particle}. 
Although percolation ideas have been proposed in the context of tortuosity in 3D porous materials \cite{ghanbarian2013percolation}, they have not been thoroughly tested near the percolation threshold.
In this contribution, we numerically show a link between the geometry of a percolating cluster and the liquid transport through porous materials with a complex geometry of pores  at the percolation threshold.

Scaling arguments from Ghanbarian and co-workers
\cite{ghanbarian2013tortuosity, ghanbarian2013percolation,hunt2017flow} suggest that the tortuosity scales, in the thermodynamics limit, with the reduced porosity according to $\tau_H \sim \delta \phi ^{\nu(1-\rm D)}$,  where $\nu$ is a critical exponent of the correlation length ($\nu \approx 0.88$ for the continuum percolation model in 3D), $\rm D$ is the fractal dimension of the cluster through which the liquid is transported, and $\delta\phi=\phi-\phi_c$. 
It was found that the fractal dimension for the most probable path through which liquid is transported is approximately $D\approx 1.43$ \cite{lee1999traveling,sheppard1999invasion,porto1997optimal,cieplak1996invasion}, implying:

\begin{equation} \label{tauresult}
\tau_H \sim \delta \phi ^{-0.38}
\end{equation}

\begin{figure*}
\includegraphics[width=\textwidth,keepaspectratio]{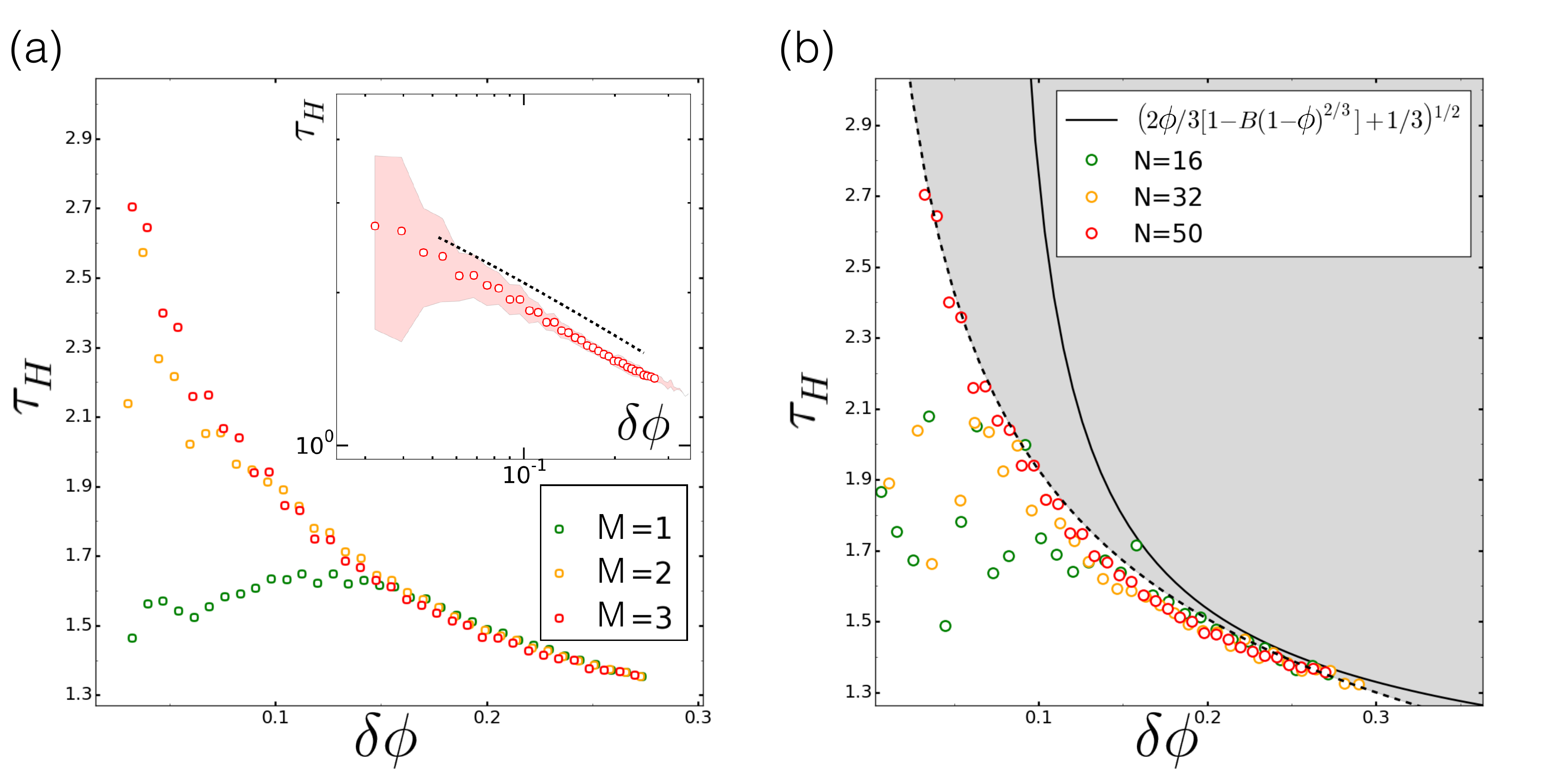}
\caption{\textbf{(a)} Hydraulic tortuoisity calculated for the system size N=50, and lattice resolution $\delta=0.04$. The fluid flow is solved on a lattice with three sizes $\tilde{\delta}=\delta/\rm M$, where M=1,2, and 3. At higher porosities, $\phi$, all three lattice refinements give similar results. Closer to the percolation threshold ($\delta \phi \lesssim 0.1$), tortuosity calculations for the liquid phase with a refinement level M=1 break down \cite{matyka2008tortuosity}. For the refinement levels M=2 and 3, the results suggest a divergence of the tortuosity at the percolation threshold ($\phi_c\approx 0.15$ for $\delta=0.04$). 
Error-bars are not given for better readability (data with error bars can be found in Fig.~\ref{fig:TauR123_error_bars}). Inset: Log-log plot of the same data. Red envelope gives one standard deviation. Dashed line with a slope $-0.38$ is given as a reference for comparison. \textbf{(b)} Hydraulic tortuoisity calculated for three different system sizes: N=16, 32, 50 and the refinement level M=3. Black dashed line is an expected tortuosity dependence $\tau_H\sim\delta\phi^{-0.38}$ in the limit of $\rm N\rightarrow \infty$. The black line provides an analytic prediction from Ahmadi {\it et al.} \cite{ahmadi2011analytical,ahmadi2012particle} with a parameter B=1.16. Details of a fitting procedure can be found in Section \ref{fitting_protocols}.}
\label{fig:TauR123}
\end{figure*}

To test this dependence, we evaluated the tortuosity from the velocity field as described in the Section \ref{lbsims}, and the results are presented in Fig.~\ref{fig:TauR123}a.
Close to the jamming threshold, $\delta\phi\approx0.25$, we find that the tortuosity is $\tau_H \approx 1.4$. This result agrees very well with experimental measurements for the packings of glass beads, $\tau_H \approx \sqrt{2}$ \cite{clennell1997tortuosity}.
For porosities close to jamming, the numerical results for all three lattice refinement levels (M=1,2,3) overlap (Fig.~\ref{fig:TauR123}a) and agree with the volume-averaged analytic prediction for mono-dispersed spheres \cite{ahmadi2011analytical,ahmadi2012particle} ({\it cf.} black line in Fig.~\ref{fig:TauR123}b).
For the porosities close to the percolation threshold, $\delta\phi\approx 0.0$, we can see that numerical simulations are consistent with the predicted divergence for the hydraulic tortuosity, Fig.~\ref{fig:TauR123}a. 
The increase of tortuosity (and its variance; inset in Fig. \ref{fig:TauR123}a) upon approaching the percolation threshold is caused by the complex geometry of the percolating cluster rather than numerical artifacts coming from the increased resolution of the liquid phase lattice, {\it cf.} Fig.~\ref{fig:threedeltas} in the Appendix.
However, divergence of a hydraulic tortuosity as $\rm\tau_H\sim\delta\phi^{-0.38}$ is expected in the thermodynamic limit, {\it i.e. $N\rightarrow\infty$}. From Equation \ref{finitetau}, we can see that for finite system sizes, where $C_I L^{-1/\nu} \gg \delta\phi$, the tortuosity is finite and reaches a maximum value at $\delta\phi=0$. This maximum tortuosity scales with the system size as $\tau_H^{\max}(N | \delta\phi=0)\sim N^{-(1-D)/d} \approx N^{0.14}$ ($D=1.43$, $d=3$, and recall that $N\sim L^d$; see Section \ref{hydraulictau} for details). In Fig.~\ref{fig:TauR123}b, we can see that the maximum tortuosity on the approach to the percolation threshold increases with the system size, and we expect that as larger systems are simulated, these values (in the limit of $\delta\phi \rightarrow 0$) will approach the scaling relation $\tau_H\sim\delta\phi^{-0.38}$, denoted by the black dashed-line in Fig.~\ref{fig:TauR123}b. In contrast to the relatively loose packings, for which lattice refinement is not crucial, lattice refinement for LB calculations is essential for the packings in the proximity of the percolation transition. This in turn sets the numerical limitations on the system size that can be feasibly simulated. A potential solution to this obstacle could be an evaluation of a geometric tortuosity \cite{clennell1997tortuosity} and leveraging on the putative relation between geometric and hydraulic tortuosities \cite{sobieski2018path}. 


\subsection{Kozeny-Carman Model of Permeability}\label{karmancozenymodel}
By construction of the Kozeny-Carman model, the liquid transport through the material is ensured down to the porosity $\phi=0$. However, this is not the case for granular porous materials. To account for this in Equation \ref{Ueq}, the porosity $\phi$ is replaced by the reduced porosity, $\phi \rightarrow \delta \phi^{\gamma} = (\phi - \phi_c)^{\gamma}$. Exponent $\gamma$ is sometimes taken {\it ad hoc} to be equal to $\gamma=1$ in references \cite{van1996network,mavko1997effect,berg2014permeability,berg2016fundamental}, however there is no firm argument supporting this particular choice. Since this exponent is yet unknown, we try to estimate $\gamma$ from a fit to the numerical data.
Knowing $\gamma$ is not crucial for highly porous materials, for which $\delta\phi \approx \phi$, but it is essential for lower porosities, where the factor $\delta\phi^{\gamma}$ contributes to the vanishing permeability $\kappa$ at the percolation threshold, $\delta\phi\rightarrow 0$.

\begin{figure}
\includegraphics[width=\columnwidth,keepaspectratio]{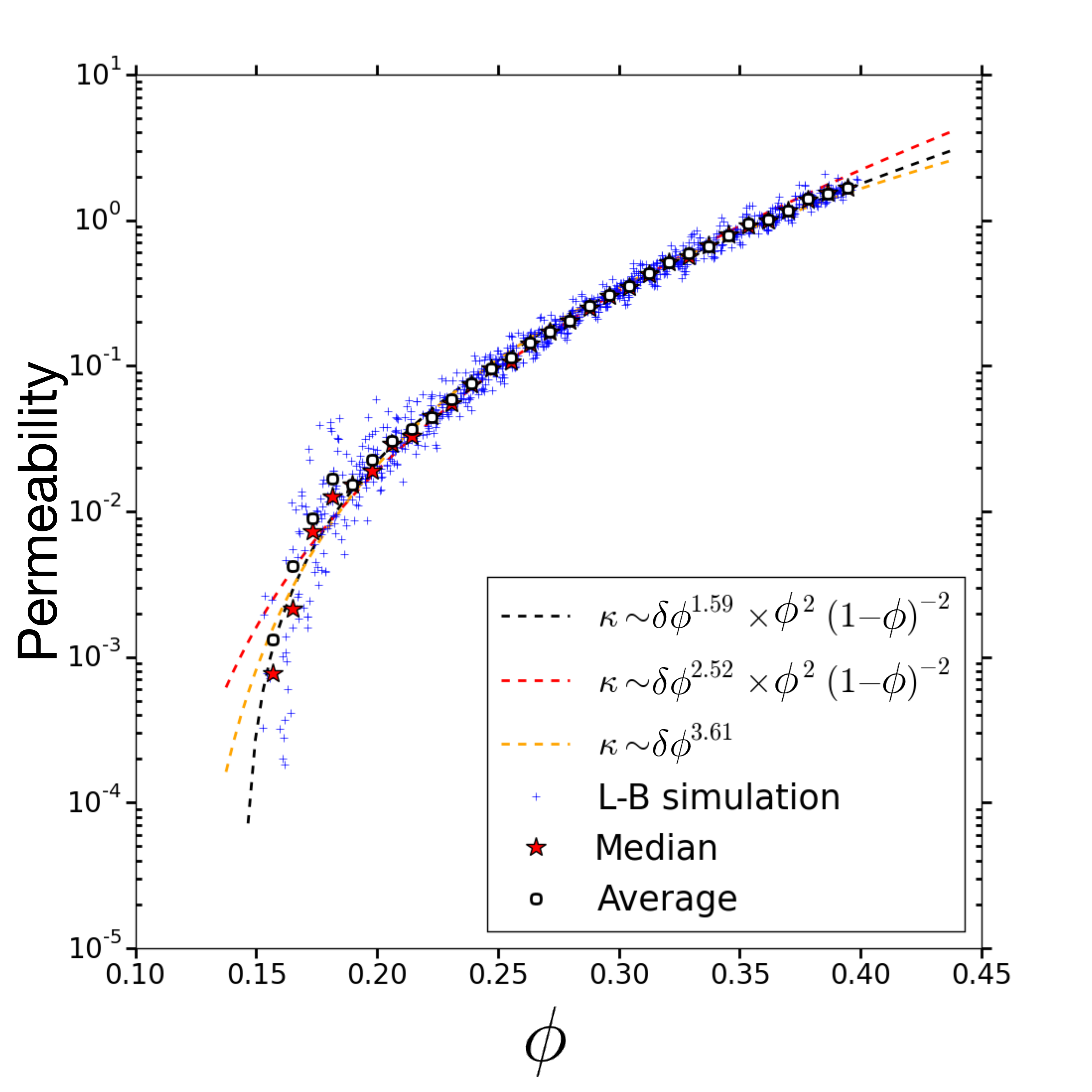}
\caption{\textbf{Permeability of deformable elastic shells packings in Darcy's regime}: Permeability obtained from Lattice-Boltzmann simulations for the system size N=50, and lattice resolutions $\delta=0.04$ and $\tilde{\delta}=\delta/3$ for the solid and fluid phases, respectively. Blue crosses represent permeability for individual simulations. Black open circles represent binned averages, and red stars correspond to medians. Permeability $\kappa$ is given in lattice units. Dashed lines correspond to three different models:  i) $\rm R_h \sim \phi/(\phi-1)$, $\rm\tau_H \sim \delta \phi^{-0.38}$, and the exponent $\gamma$ being a fitting parameter; ii) $\rm R_h \sim \phi/(\phi-1)$, $\gamma = 1.76$, and $\tau_H \sim \delta \phi^{-0.38}$; iii) power-law {\it ansatz}: $\kappa\sim\delta\phi^{\bar{e}}$. Fitting details can be found in Section \ref{fitting_protocols}.}
\label{fig:kappa}
\end{figure}

In Section \ref{percothreshold}, we found numerically that the percolation threshold depends on the resolution of the used lattice.
Moreover, in Section \ref{rhresults} we found that the hydraulic radius reaches 0 at the porosity $\phi=0$, and does not strongly depend on the lattices resolution $\delta$.
Finally, in Section \ref{taudivergence} we found that the tortuosity of flow stream lines diverges upon the approach of the percolation threshold, consistent with the prediction $\tau_H \sim \delta \phi^{-0.38}$.
Using Equations \ref{eq:kappa}, \ref{rinvresult}, and \ref{tauresult}, we can put together a relationship between material porosity and permeability $\kappa$ that reads:
\begin{equation} \label{kappafinal}
\kappa = \rm C_{\kappa} \times \frac{\delta \phi^{\gamma+0.76}\phi^2} {\left(1-\phi\right)^2}
\end{equation}
where $\rm C_{\kappa}$ is a constant.
A fit of this model is presented in Fig.~\ref{fig:kappa} (black dashed-line). Results are given for the lattice resolution $\delta =0.04$, for which the tortuosity diverges and the flow ceases at porosity $\phi_c\approx 0.15$. We can see in Fig.~\ref{fig:kappa} that Equation~\ref{kappafinal} captures quite accurately the change of the material permeability $\kappa$ in a broad range of porosities \textemdash~from the onset of the jamming up to the percolation threshold, and regardless of the model fitting method, {\it cf.} Fig.~\ref{fig:kappa} and Fig.~\ref{fig:manyfits}. Depending on the fitting procedure, the value of the exponent $\gamma$ varies slightly, with the average (over four different fitting procedures) value $\gamma = 0.89\pm 0.15$. This is quite close to the value used {\it ad hoc}, $\gamma=1.0$.
In the limit of the large porosities, {\it i.e.} where  $\phi \gg \phi_c^*$, we can approximate $\delta \phi^{\gamma} \approx \phi^{\gamma}$, which reduces Equation \ref{kappafinal} to a simpler form $\kappa \sim \phi^{3.59}/(1-\phi)^2$ (with $\gamma \approx 0.83$). Interestingly, this approximate form, with a fractional power close to 3.6, is in good agreement with recent experimental and numerical work, where this exponent was estimated to be 3.7 (for porosities such that $\phi - \phi_c \approx \phi$) \cite{chen2008pore,chen2009temporal}.
It is worth noting that although the above model depends on a value of $\phi_c$ (which also encompasses finite-size effects), it does not affect the generality of the model because of two reasons: i) the value of the hydraulic radius is quite insensitive to the lattice size used in the calculations; ii) flow tortuosity and dilution of the capillaries is determined by a reduced porosity $\delta\phi$, thus Equation \ref{kappafinal} should apply for various system and lattice sizes in the vicinity of the percolation threshold even though the exact percolation thresholds are different.

In this work we compare Equation \ref{kappafinal} to a {\it scaling ansatz}  $\kappa\sim\delta\phi^{\bar{e}}$, a good guess for the transport properties in disordered systems and close to the critical point \cite{martys1994universal}. Halperin {\it et al.} \cite{halperin1985differences,halperin1987transport} showed that there are several universality classes of porous media where the scaling exponent $\bar{e}$ depends on the model's details. For example, in the so called the {\it Swiss-cheese} model, $\bar{e}\approx 4.4$, whereas for the Inverted {\it Swiss-cheese} model, $\bar{e}\approx 2.4$. The relation $\kappa\sim\delta\phi^{\bar{e}}$ fits the data in a broad range of porosities, yellow dashed-lines in Fig.~\ref{fig:kappa} and Fig. \ref{fig:manyfits}.
However, the fitted exponent values depend on the fitting procedure and vary in the range of $\left[2.72,3.88\right]$, with an average value $\bar{e}=3.4$. Moreover, the estimated percolation threshold ($\phi_c$) differs noticeably from the estimations made in Fig.~\ref{fig:perctrans}c. 
Despite the fact that the power-law scalings are often very useful, it is not always clear how the scaling exponents relate to the connectedness of the pores and the tortuosity of the flow \cite{martys1994universal}.
Additionally, in Fig.~\ref{fig:manyfits}a and Fig.~\ref{fig:manyfits}b,  we compare our numerical data to the standard Kozeny-Carman model, where $\kappa\sim\phi^3/(1-\phi)^2$ \cite{carman1937fluid,macdonald1991generalized,mavko1997effect,xu2008developing,yu2008analysis,bear2010}.
This classical model has been successfully applied to many porous materials for which $\phi-\phi_c \approx \phi$ \cite{mcgregor1965effects,macdonald1991generalized,bear2010}.
Some authors extended the Karman-Cozeny model by accounting for fractal geometry of porous materials \cite{costa2006permeability,xu2008developing,yu2008analysis}, but these models still assume permeability down to porosity $\phi=0$.
However, in this work, we study permeabilities in the range of the porosities for which the above approximation does not hold. Therefore, the classical Kozeny-Carman model performs worse, as shown by the green dashed-line in Fig.~\ref{fig:manyfits}a and Fig.~\ref{fig:manyfits}b.


\section{Discussion \& Conclusions}
Our results support a simple model of the fluid flow retardation in deformable granular materials, compressed from the onset of mechanical stability at the jamming point down to the percolation threshold. Porous material is essentially described as a collection of tortuous and randomly placed capillaries, where, close to the percolation threshold, tortuosity and capillaries dilution dominate liquid transport.
We have shown that upon compaction, the void space between pressurized elastic shells undergoes a sharp system-size dependent transition. We also find that the hydraulic radius vanishes in a lattice-resolution independent manner as the porosity diminishes. Next, using Lattice-Boltzmann simulations, we have shown that tortuosity of the flow stream lines abruptly increases at the percolation threshold.
In Equation \ref{eq:kappa}, the effects of the capillaries' density and tortuosity are factorized, and this has motivated a substantial research devoted to tortuosity \cite{clennell1997tortuosity,ghanbarian2013tortuosity}.
Combined with a percolation scaling theory, we were able to support the fractional dependence of tortuosity on the porosity of the sample. Our work underscores that at higher porosities, where the fluid flow is not tortuous ($\tau_H$ is mildly varying for larger $\phi$), the major geometric determinant of the flow obstruction is the amount of the void space accessible to fluid \textemdash~captured in the quadratic dependence on a hydraulic radius $R_h$. In turn, upon the approach to the percolation threshold, the complex geometry of liquid transporting channels ultimately leads to flow hindrance. Nonetheless, the dilution of the capillaries upon the approach to the percolation threshold, described by the $\gamma$ exponent, remains elusive. We found numerically that $\gamma\approx 0.89\pm0.15$, which is close to the {\it ad hoc} value $\gamma=1.0$ \cite{van1996network,mavko1997effect,berg2014permeability,berg2016fundamental}, but this value does not have a firm grounding in the percolation theory. In Section \ref{percogamma}, we present a simple scaling argument from the percolation theory that suggests this exponent to be $\gamma=1.76$. If tortuosity is neglected, this would explain our numerical data very well. However, when the tortuosity contribution is included, this leads to the decay of the permeability in the vicinity of the percolation threshold with the exponent close to 2.5, {\it i.e.} $\kappa\sim\delta\phi^{2.52}$. Despite the fact that this is close to the Inverted {\it Swiss-cheese} model exponent ($\bar{e}\approx 2.4$), it does not reproduce the numeral data well, {\it cf.} Fig.~\ref{fig:kappa} and Fig.~\ref{fig:manyfits}. However, it is worth noting that the scaling argument given for $\gamma$ (see Section \ref{percogamma}) is a geometric one, whereas the liquid transport is a dynamic process, and the number of hypothetical capillaries may differ from the number of possible percolating paths. Additionally, the dynamic universality class for transport properties splits for lattice and continuum percolation \cite{halperin1985differences,spanner2015splitting}, therefore drawing conclusions from the numerical calculations performed in a discretized domain close to the critical point requires caution.
This intriguing results motivate further research on the capillary model in the proximity of the percolation threshold within a framework of the percolation theory. Additionally, this work, alongside the works of others \cite{arns2005cross,coleman2008transport,berg2012archie,ghanbarian2013tortuosity,berg2016fundamental}, can be potentially useful in studying other transport processes like, for example, electrical conductivity of an electrolyte (as well as the electrical tortuosity $\tau_e$ \textemdash~an analog of the hydraulic tortuosity $\tau_{\textrm{H}}$ in the fluid transport) \cite{archie1942electrical,clennell1997tortuosity,wong1999conductivity,ghanbarian2013tortuosity}.

Finally, in our work we considered only packings of identical shells. In Section \ref{appx:hydraulicradius} we can see that poly-dispersity seems to contribute only a constant factor in the relation for $\rm R_h$, Equation \ref{finalRh}, without changing its functional dependence on the porosity $\phi$. Furthermore, in 3D packings of unequal spheres, polydispersity has only a minor impact on the percolating clusters \cite{van1996network, rintoul2000precise}. Therefore, Equation \ref{kappafinal} may be applicable to other disordered and compacted systems made of deformable particles.

\section{Acknowledgments}
We thank Jayson Paulose for feedback on the manuscript. We also thank two anonymous reviewers for their insightful comments and suggestions. This research used resources of the National Energy Research Scientific Computing Center, a DOE Office of Science User Facility supported by the Office of Science of the U.S. Department of Energy under Contract No. DE-AC02-05CH11231.
This work was supported by a grant from the Simons Foundation
(\#327934, O.H.), by a NSF Career Award (\#1555330, O.H.) and a NIH grant (R01GM115851).

\section{APPENDIX}
\subsection{Source Code Availability}
The source code for the Lattice-Boltzmann calculations is available on GitHub \cite{sourcecode1}.

\subsection{Generation of Jammed Packings}\label{appx:jammig}
To generate jammed packings, we randomly place particles in a cubic box with periodic boundary conditions.
The initial radii of these spherical particles are set such that the initial volume fraction is about $\psi_0=0.01$. Next, we successively increase or decrease the radii of the particles,
with every change followed by the energy minimization with the FIRE algorithm \cite{bitzek2006structural} and velocity-verlet integrator \cite{allen1989computer}. The parameters used in the FIRE algorithm are: $\rm dt_{\rm FIRE}=0.1$, $\rm dt_{\rm FIRE}^{\max}=1.5$, $\alpha_{\rm FIRE}=0.1$, $\rm N_{min}=5$, $\rm f_{\alpha}=0.99$, $\rm f_{\rm inc} = 1.1$, $\rm f_{\rm dec}=0.25$. The termination condition for the FIRE algorithm is: $\underset{i}{\max} ~ | \mathbf{f}_{\rm i} | \le 10^{-15}$.

Initially, for each inflation step, the particle's radius is increased following the rule: $\rm r_{\rm new}=\rm r_{\rm old}\cdot\left(1+\epsilon_r\right)$, where initially, $\epsilon_r=0.01$. The forces between particles are Hertzian: $\mathbf{F}(\mathbf R)=-\frac{4}{3}E^*\sqrt{R^*}\hat{\mathbf{R}}\rm h^{3/2}$, where $\rm h$ is an overlap between particles, $\hat{\mathbf{R}}$ is a unit vector along $\mathbf{R}$, $\rm E^*=\rm E/2(1-\nu^2)$ is an effective Young's modulus, $\rm R^*=\rm r/2$ is an effective radius, and $\rm r$ is the radius of a particle. In this work, we use $\rm E=1$, $\nu=0.5$. 
The pressure in the simulation box is calculated as: $ P = -\frac{1}{3} \sum_{\alpha} \sigma_{\alpha\alpha}$, where the stress tensor $\sigma_{\alpha\beta}$ is obtained from the virial formula:
$ \sigma_{\alpha \beta} =
 - \frac{1}{\mbox{V}}   \sum_{i} \sum_{i > j} \mathbf{r}^{\alpha}_{\rm ij} \mathbf{F}^{\beta}_{\rm ij}$, where $\mathbf{r}^{\alpha}_{\rm ij}$ is the $\alpha^{\rm th}$ component of the vector pointing from the center of a particle j to i, and $\mathbf{F}^{\beta}_{\rm ij}$ is the $\beta^{\rm th}$ component of the contact force between particles i and j.
 
\begin{figure}
\includegraphics[width=\columnwidth,keepaspectratio]{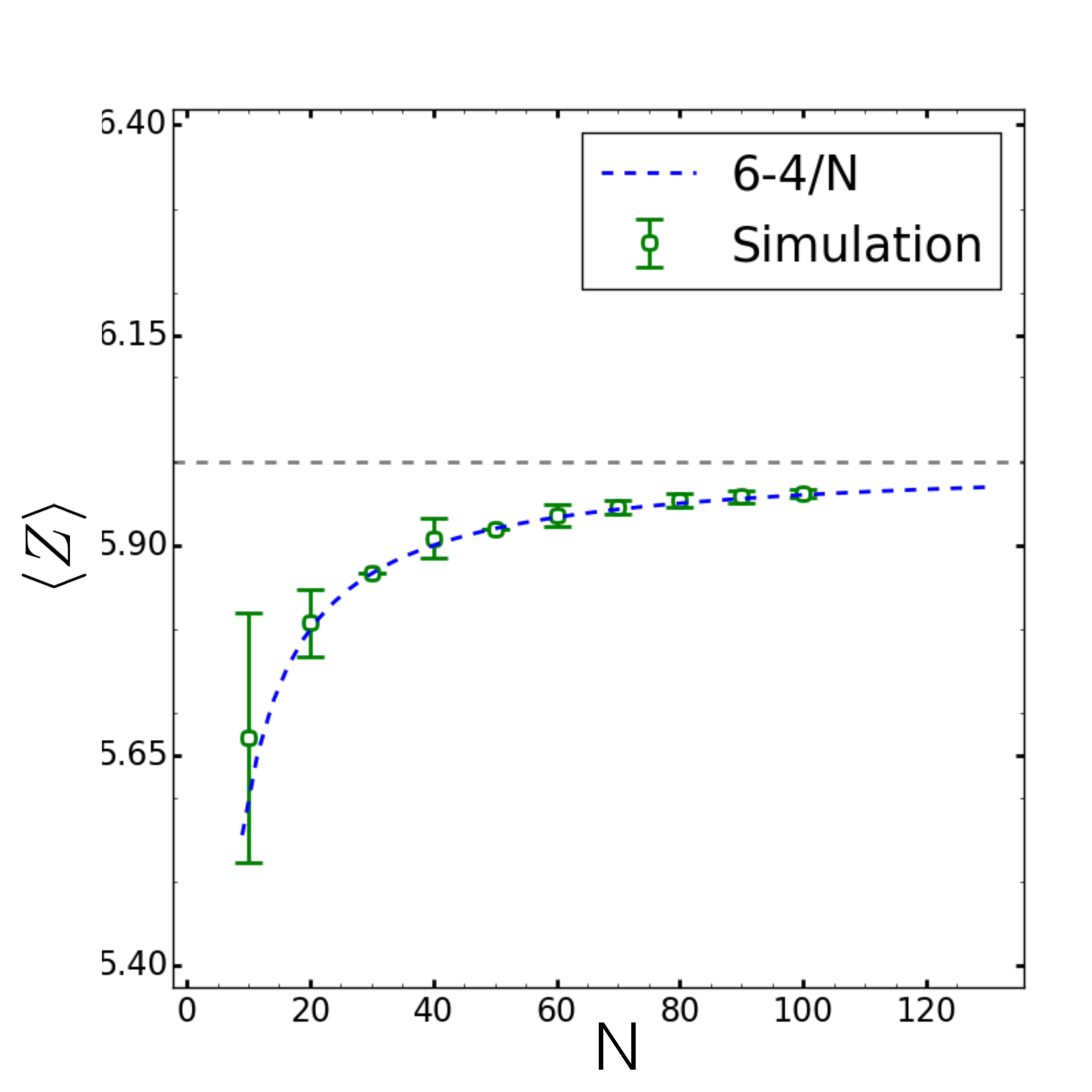}
\caption{\textbf{Average contact number}: A mechanically stable system must have a force balance on each particle. For N spheres in d dimensions, the number of constraints that has to be satisfied by the inter-particle forces is $\rm d \times \rm N$. In the system with periodic boundaries this number is $\rm d \times \rm N - \rm d$. Additionally, there is one more degree of freedom \textemdash~a volume fraction at the jamming \textemdash~that has to be constrained. Thus, the counting argument provides the number of constraining equations that needs to be $\rm N_c = \rm d \times \rm N - \rm d +1$. According to Maxwell's criterion, the number of inter-particle contacts, $\rm N \left <\rm Z\right>/2$, must be at least equal to the number of equations $\rm N_c$. For frictionless spheres the packing at the jamming point has exactly this number of contacts: $\left <\rm Z\right> = 2 \rm d - 2(\rm d-1)/N$; which, for 3D, the average number of contact per particle is $\left <\rm Z\right> = 6 - 4/N$ \cite{o2003jamming}. The results are in the range of N that is meaningful for the present work: (10,100).
For each N, 100 different packings are generated. Error bars give one standard deviation. }
\label{fig:ZvsN}
\end{figure}

\begin{figure}
\includegraphics[width=\columnwidth,keepaspectratio]{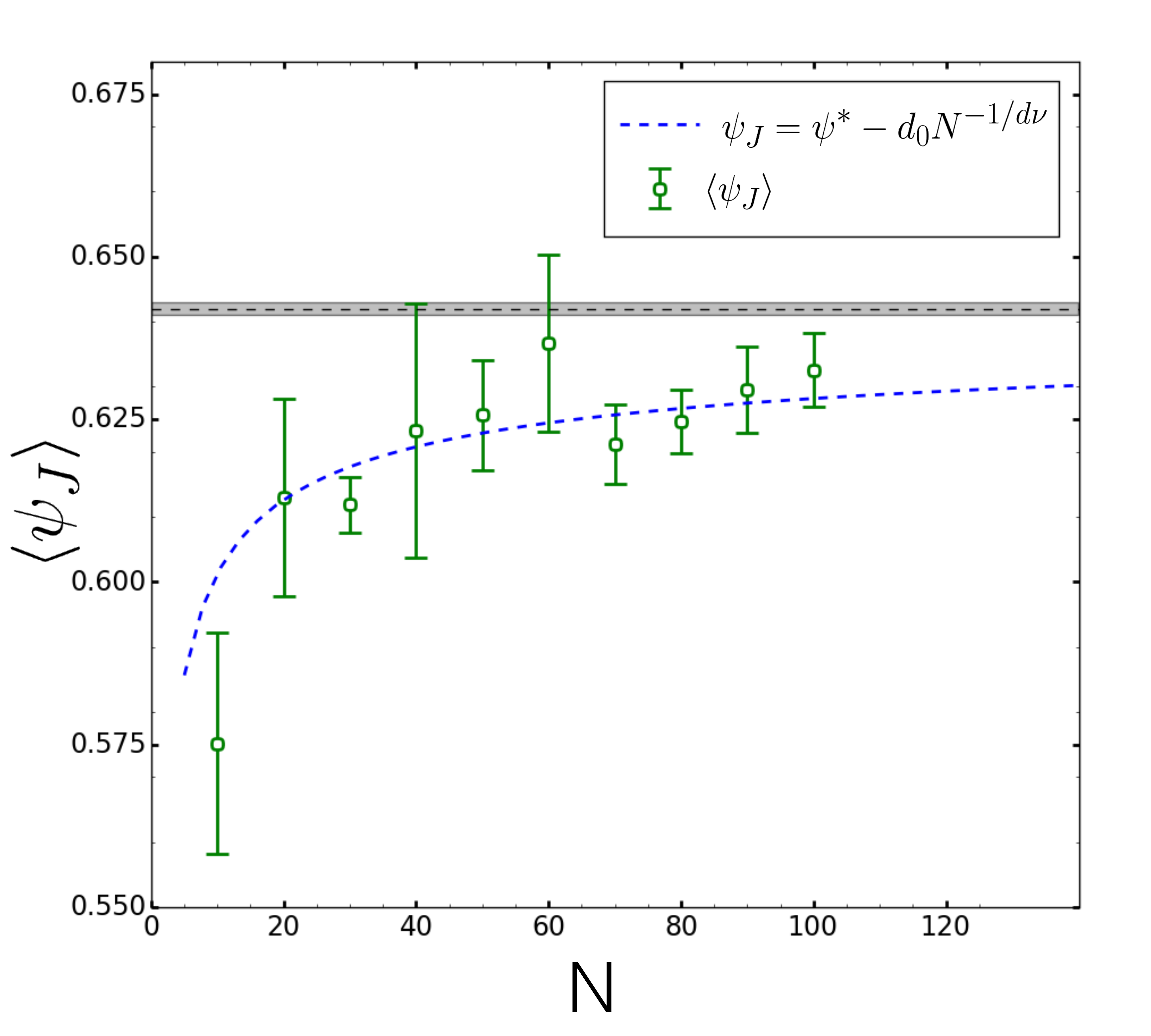}
\caption{\textbf{Distribution of the volume fraction at the jamming threshold $\psi_J$}: The position of the maximum of the jamming volume fraction distribution exhibits finite-size scaling: $\psi^* - \psi_J = d_0 N^{-1/d\nu}$, where $d_0=0.12 \pm 0.03$, $d=3$, $\nu=0.71\pm 0.08$, and $\psi^* =0.639 \pm 0.001$ \cite{o2003jamming}. The asymptotic value is plotted as a shaded area. The green dots in the plot are the average values calculated from 100 independent simulation. Error bars give one standard deviation.}
\label{fig:PhivsN}
\end{figure}

When the pressure of the packing is greater than $\rm P > 2\cdot \rm P_{\min}=2\cdot 10^{-8}$, the parameter $\epsilon_r$ is halved, and the particles' sizes are deflated according the rule: $\rm r_{\rm new}=r_{\rm old}\cdot\left(1-\epsilon_r\right)$. When the pressure drops below $\rm P_{\min}=10^{-8}$, then $\epsilon_r$ is again halved and the particles are inflated. 
The process continues until the pressure $\rm P$ settles at the value $\rm P_{\min} < \rm P < \rm 2 P_{\min}$. If the packing contains any rattler, the configuration is rejected and the procedure is repeated. 
The final configuration provides positions of soft-spheres particles that are next replaced by the Finite Element representation.
The packings generated using the described algorithm have been tested in terms of the number of contacts (Fig. \ref{fig:ZvsN}) and the finite size effects on the volume fraction at the jamming point (Fig. \ref{fig:PhivsN}) \cite{o2003jamming, gniewekphdthesis}.

\subsection{Hydraulic Radius: Geometric Argument \cite{ng1986model,macdonald1991generalized}}  \label{appx:hydraulicradius}
For a packed bed of spherical particles with a particle size distribution $\rm n(\rm D_p)$, the $\rm i^{\rm th}$ moment of the particle size distribution is:

\begin{equation}\label{eq:momenti}
\rm \mu_i = \int_{0}^{\infty} \rm D^i_p \rm n(\rm D_p) d\rm D_p
\end{equation}
\noindent
If a horizontal cut is made across the packing, one obtains circular disks of the size $\rm x$, projected on the sectional plane (this assumption holds only approximately for more compact systems). The size distribution of these disks is:
\begin{equation}\label{eq:fx}
\rm f (x) = \int_0^{\infty} P(x|D_p)P(D_p)d\rm D_p
\end{equation}
\noindent
Here $\rm P(D_p)$ is the pdf of $\rm D_p$:

\begin{equation}
\rm P(D\rm _p) = \frac{\rm n(\rm D_p)}{\int n(D_p)d \rm D_p} = \frac{\rm n(\rm D_p)}{\rm \mu_0}
\end{equation}
\noindent
$\rm P(x|\rm D_p)$ is a conditional probability density function that given a sphere diameter $\rm D_p$
 the diameter of a given disc in a plane cut ranges between $\rm x$ and $\rm x + d\rm x$. Note that disks of the same size can originate from spheres of a different size because the disc  size depends on the position at which a sphere is cut.
 
It has been shown that a plane cut through a random spheres packing provides a distribution of disks on a plane that follows \cite{ng1986model,macdonald1991generalized}:
\begin{equation}\label{eq:pxd}
\rm P(x|\rm D_p) = \frac{x}{\rm D_p\sqrt{\rm D_p^2-\rm x^2}} \left [ 1 - \Theta(\rm x - \rm D_p)\right]
\end{equation}
\noindent
where $\Theta(\cdot)$ is the Heaviside function. 
Substituting Equation \ref{eq:pxd} into Equation \ref{eq:fx}, we get:
\begin{equation}
\rm f (x) = \int_0^{\infty} \frac{\rm n(\rm D_p)}{\mu_0}\frac{x}{\rm D_p\sqrt{\rm D_p^2-\rm x^2}} \left [ 1 - \Theta(\rm x - \rm D_p)\right] d\rm D_p
\end{equation}
\noindent
Thus, for a given plane cut, the amount of the surface occupied by the disks on that plane is given as:
\begin{equation}
\alpha = N_{\alpha} \frac{\pi}{4}\int_0^{\infty} \rm x^2 f(\rm x)d\rm x = N_{\alpha} \frac{\pi}{6}\frac{\rm \mu_2}{\rm \mu_0}
\end{equation}
\noindent
where $\rm N_{\alpha}$ is the number of discs per unit cross-section area. Integrating over the whole body, we obtain the volume of the solid material: $\rm V=\alpha L^3$, where $\rm L$ is a linear dimension of a body. We can see that $\alpha$ is proportional to the volume fraction $\psi = \rm V/\rm L^3$, and finally $\rm N_{\alpha} \propto \psi = 1 - \phi$, where $\phi$ is the material's porosity. 
Similarly, the wetted perimeter per unit area of bed $\Sigma$ can be obtained from:
\begin{equation}
\Sigma = \rm N_{\alpha} \pi \int_0^{\infty} \rm x \rm f(\rm x) d\rm x = \rm N_{\alpha} \frac{\pi^2}{4}\frac{\rm \mu_1}{\rm \mu_0}
\end{equation}
\noindent
leading to $\Sigma \sim \rm N_{\alpha} \propto 1-\phi$.

\noindent
Finally, the hydraulic radius $\rm R_h$ is:

\begin{equation}\label{finalRh}
\rm R_h = \frac{1-\alpha}{\Sigma} = \frac{2}{3\pi}\frac{\phi}{1-\phi}\frac{\rm \mu_2}{\rm \mu_1}
\end{equation}
\\
\noindent
For the monodisperse spheres packings, the size distribution in Equation \ref{eq:momenti} is given by the Dirac delta function $n(D_p)\equiv\delta(D_p-D_0)$. This leads to $\mu_2=D_0^2$ and $\mu_1=D_0$, which reduces Equation \ref{finalRh} to $R_h/D_0=\frac{2}{3\pi}\frac{\phi}{1-\phi}$.

\subsection{Tortuosity Calculation} \label{appx:tau_calc}
For the fluid flow, hydraulic tortuosity $\tau_H$ is defined as:
\begin{equation}\label{taudef}
\tau_H = \frac{\left<\lambda\right>} {L } \geq 1
\end{equation}
where $\left<\lambda\right>$ is the mean length of the fluid particles paths and L is a linear dimension of a porous medium in the direction of a macroscopic flow. Despite this simple definition, tortuosity is not easy to measure experimentally and computationally. 
In real porous media, flow streams are complicated, as the fluid fluxes continuously change their sectional area, shape, and orientation, or the flow streams branch and rejoin. It is also not clear how the average in Equation \ref{taudef} should be calculated: over the whole volume or over the planar cross-section, and if so, what is the most proper cross-section to do this? 
It has been concluded that the proper hydraulic tortuosity should be calculated as an average in which streamlines are weighted with fluid fluxes \cite{clennell1997tortuosity, duda2011hydraulic, matyka2012calculate}.
Thus tortuosity can be calculated as:
\begin{equation}
\tau_H = \frac{\sum_i \tilde{\lambda}_i \omega_i}{\sum_i \omega_i}
\end{equation}
where $i$ enumerates discrete streamlines, $\tilde{\lambda}_i = \lambda_i /L$, $\lambda_i$ is the length of the $i^{th}$ streamline with the weight $\omega_i = 1 /t_i$, where $t_i$ is a time in which fluid particles move along the $i^{th}$ streamline \cite{duda2011hydraulic}. The rationale behind the $\omega_i$ factor is to weigh each streamline proportionally to the volumetric flow associated with a streamline.
For the incompressible flow, $t_i$ tells how long it takes for the particles in a given streamline to travel a distance $L$ in a macroscopic flow direction. Thus, the average component of the velocity for that streamline, in a direction of the flow, is proportional to the weight factor $\left< v_x\right>_{i} \sim \omega_i$.
Extending this idea in the continuous limit, for a cross-section perpendicular to the macroscopic flow, the hydraulic tortuosity can be formulated as:
\begin{equation}\label{eq:gentau}
\tau_H = \frac{ \int_A u_{x}(\mathbf r) \tilde{\lambda}(\mathbf r) d\mathbf\sigma}{\int_A u_{x}(\mathbf r)d\mathbf\sigma}
\end{equation}
\noindent
where A is a cross-section perpendicular to the axis x, both integrals are taken over the surface $d\mathbf\sigma \in A$, $\tilde{\lambda}(\mathbf r)$ is the length of a streamline intersecting with the surface A at the location $\mathbf r$ (normalized by $L$), and $u_x(\mathbf r)$ is the component of the velocity field at $\mathbf r \in A$ normal to A. Moreover, it was shown that the cut can be done not necessarily in a direction of the macroscopic flow but in principle in any direction \cite{duda2011hydraulic}. 
Even though there is freedom in the location of where the cut can be done, both integrals are still difficult to calculate numerically \cite{matyka2008tortuosity}.

This numerical problem can be bypassed by noticing that \cite{duda2011hydraulic}:
\begin{equation}\label{eq:gentau2}
\tau_H = \frac{ \int_A u_{\bot}(\mathbf r) \tilde{\lambda}(\mathbf r) d\sigma}{\int_A u_{\bot}(\mathbf r)d\sigma} = \frac{ \int_V u(\mathbf r) d\nu}{\int_V u_x(\mathbf r)d\nu}
\end{equation}
\noindent
and the r.h.s. can be further simplified as \cite{duda2011hydraulic}:
\begin{equation}\label{eq:simpletau1}
\tau_H = \frac{\left<u\right>}{\left<u_{x}\right>}
\end{equation}
\noindent
This form of tortuosity is particularly handy in numerical analysis since it requires only solving the flow field without struggling with resolving streamlines \cite{matyka2008tortuosity,matyka2012calculate}.
Some inaccuracies may occur in Equation \ref{eq:simpletau1} if eddies exist in the flow. Although it cannot be assured that such structures do not occur in complex porous materials, the contribution from eddies to Equation \ref{eq:gentau} is negligible at low Reynolds numbers \cite{duda2011hydraulic}.

Finally, the velocity field is found with Lattice-Boltzmann simulations. Then, $\tau_H$ can be calculated from the values of the flow at each node in the lattice:
\begin{equation}\label{eq:simpletau2}
\tau_H = \frac{\sum_{\mathbf r} u(\mathbf r) }{\sum_{\mathbf r} u_x(\mathbf r) }
\end{equation}
where $\mathbf r$ runs over all lattice nodes \cite{matyka2012calculate}.

\subsection{Hydraulic Tortuosity: Percolation Theory Argument}
\label{hydraulictau}
The evolution of the void region between overlapping, randomly located spheres undergoes a percolation transition \cite{van1996network,priour2014percolation}. This transition exhibits a critical behavior and falls into a continuum percolation universality class \cite{van1996network,rintoul1997precise,priour2014percolation}.
For porous materials, a porosity $\phi$ acts like the percolation probability in a classical percolation theory.
Above a certain porosity threshold $\phi_c$, there exists a cluster that spans the whole system and facilitates fluid transport. This has been leveraged to connect tortuosity with material porosity \cite{ghanbarian2013tortuosity, ghanbarian2013percolation,hunt2017flow}. Here we present an equivalent but simpler argument.
\\

\noindent
Percolation theory predicts that a mean distance $\xi$ between any two sites on a cluster is given by a scaling law \cite{SA94}:
\begin{equation}
\xi \sim |\rm \phi - \rm \phi_c|^{-\nu}
\end{equation}
where $\nu$ is a critical exponent of the correlation length.
The total length of a walk $\lambda$ constructed on that cluster has a fractal dimension $D$ and reads $\lambda \sim \xi^D$ \cite{havlin1987diffusion}. At the percolation threshold, the correlation length $\xi$ diverges and is the same as the system size. From the definition of a tortuosity $\rm\tau_H$, we have then (close to the percolation threshold and $\phi > \phi_c$):
\begin{equation}
\rm\tau_H = \frac{\lambda}{\xi} \sim \xi^{D-1} \sim |\phi - \phi_c|^{\nu(1-D)} \equiv \delta \phi^{\nu(1-D)}
\end{equation}

\noindent
For a finite system, there is an additional finite-size correction that accounts for the shift of the percolation transition. Taking this into account, the scaling for $\rm\tau_H$ reads:
\begin{equation}\label{finitetau}
\tau_H \sim |\phi - \phi_c + C_{I}\cdot L^{-1/\nu}|^{\nu(1-D)}
\end{equation}
where $C_{I}$ is a constant and it is of the order of $C_{I}\sim \mathcal{O}(1)$.
\\
It has been shown that the most probable traveling length of an incompressible flow on a percolating cluster falls into the same universality class as the optimal path in strongly disordered media and the shortest path in the invasion percolation with trapping \cite{lee1999traveling,sheppard1999invasion}~\textemdash~for which the fractal dimension is $D \approx 1.43$ \cite{porto1997optimal,cieplak1996invasion}.
Finally, taking the exponent $\nu \approx 0.88$, one gets a scaling law for  tortuosity ($\rm L \rightarrow \infty$): $\tau_H \sim |\phi - \phi_c|^{-0.38} \equiv \delta \phi^{-0.38}$. For finite systems ($\rm N < \infty$), tortuosity reaches maximum value at $\delta\phi=0$, which scales with the system size as $\tau_H^{\max}\sim L^{-(1-D)} = N^{-(1-D)/d} \approx N^{0.14}$, where $d=3$ is a system dimension.

A similar scaling argument was numerically tested for 2D overlapping squares on a Cartesian lattice \cite{duda2011hydraulic}, where via finite-size scaling analysis, it was shown that the tortuosity in the neighborhood of percolation transition is controlled by the fractal geometry of a percolating channel.

\subsection{Scaling Argument for the $\gamma$ Exponent}\label{percogamma}
Taking a planar cut through the porous material, we observe $n_c$ capillaries distributed over the area of the cut. If the material is isotropic, the direction of the cut does not matter, and we can assume that the cut is made perpendicularly to the direction of fluid transport. This plane-cut would obviously contain cross-sections of all the capillaries that are responsible for the liquid transport thorough the material in the given direction. 
Close to the percolation threshold, we expect to have a single capillary in the area that is proportional to $\xi^2$, where $\xi$ is the correlation length. If that is the case, the expected number of capillaries penetrating thorough the material is $\rm n_c \propto L^2/\xi^2$, where $\rm L$ is the linear size of the body.
$\xi$ is related to the exponent of the correlation length ($\nu\approx 0.88$) as $\xi\sim \delta\phi^{-\nu}$. Therefore, we have a power-law relation between the number of capillaries and $\delta\phi$ which reads $\rm n_c\sim\delta\phi^{2\nu} \approx \delta\phi^{1.76}$. 

\subsection{Parameters fitting procedure}\label{fitting_protocols}
Parameters fitting and standard deviation estimations are done with a non-linear least squares method from the \texttt{scipy} Python library.

\subsubsection{Extrapolating percolation threshold to the continuum limit}\label{perc_thr}
In Fig.~\ref{fig:perctrans}d, we extrapolate a percolation threshold down to the continuum limit $\phi_c^*$, {\it i.e.} $\delta\rightarrow 0$. To that end, we fit a sigmoid function to the percolation probability data in Fig.~\ref{fig:perctrans}c. Next, for different $\delta$ we take a porosity at which the percolation probability is equal to $1/2$ as a percolation threshold. Finally, we fit a power-law dependence: $\phi_c - \phi_c^* = C_N\cdot \delta^{\beta}$. The fitting results are in Table~\ref{tab:fits} (row: Fig.~\ref{fig:perctrans}d). Parameters are obtained as a result of minimization of the function: $\rm Error\propto\sum_i\left(\phi_{c,i}^{num}-\phi_{c,i}^{fit}\right)^2$, where $\rm\phi_{c,i}^{num}$ is a percolation threshold estimated from the numerical data, and $\rm \phi_{c,i}^{fit}$ is estimated from the power-law dependence for varying $\phi_c^*$, $C_N$, and $\beta$.

\subsubsection{Fitting power-law dependences for tortuosity}
We fit a power-law dependency for tortuosity data obtained from Lattice-Boltzmann simulations. The relation has a functional form $\rm\tau_H=C_{\tau}(\phi-\phi_c)^{-0.38}$, where there are only two fitting parameters: $\phi_c$ and a constant factor $\rm C_{\tau}$. Porosity $\phi$ is a value known from Finite Elements simulations, and the exponent $-0.38$ is predicted from a percolation theory, see Section \ref{hydraulictau}. 
We perform a non-linear fit by minimizing the error function: $\rm Error\propto\sum_i\left(\tau_{i}^{num}-\tau_{i}^{fit}\right)^2$, where the index $i$ runs over all experimental samples, $\rm\tau_i^{num}$ is a numerical tortuosity from LB simulations for the system $i$, whereas $\rm\tau_i^{fit}$ is a fit to the power-law dependency. The results are given in the Table~\ref{tab:fits} (row: Fig.~\ref{fig:TauR123}b).

\subsubsection{Parameters estimation for permeability}
\begin{figure*}[ht]
\includegraphics[width=0.9\textwidth,keepaspectratio]
{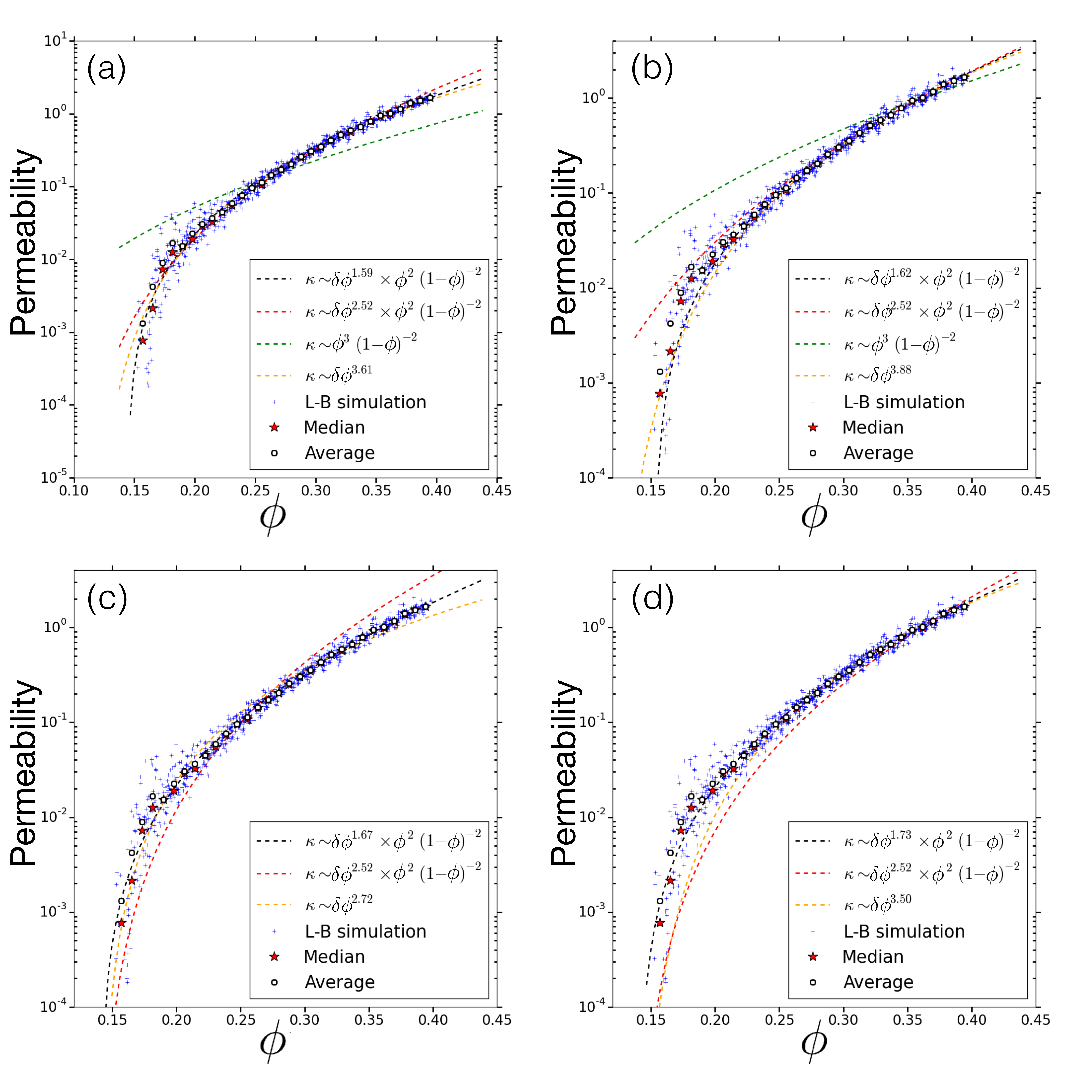}
\caption{Permeability obtained from Lattice-Boltzmann simulations for the system size N=50, the lattice resolution $\delta=0.04$, and the lattice size for the fluid phase $\tilde{\delta}=\delta/3$. Symbols are the same as in Fig.~\ref{fig:kappa}; blue crosses represent permeability for individual simulations, black open circles represent binned averages, red stars are median values, and dashed lines correspond to different models. Permeability $\kappa$ is given in lattice units. Fitted parameters are given in Table \ref{tab:fits}. In Fig.~\ref{fig:manyfits}a and Fig.~\ref{fig:manyfits}b, $\phi_c$ is a free parameter, whereas in Fig.~\ref{fig:manyfits}d, $\phi_c$ is fixed at $\phi_c=0.141$ (see Section \ref{perc_thr}). Classical Kozeny-Carman model ($\kappa\sim\phi^3/(1-\phi)^2$) is given for a comparison in Fig.~\ref{fig:manyfits}a and Fig.~\ref{fig:manyfits}b. Error functions used in a fitting procedure are given in Table~\ref{tab:fits}.}
\label{fig:manyfits}
\end{figure*}

Fits are done for three different permeability relations: i) $\kappa=C_{\kappa}(\phi-\phi_c)^{\gamma+0.76}\phi^2(1-\phi)^{-2}$, ii) $\kappa=C_{\kappa}(\phi-\phi_c)^{2.52}\phi^2(1-\phi)^{-2}$, and iii) $\kappa=C_{\kappa}(\phi-\phi_c)^{\bar{e}}$. In Fig.~\ref{fig:kappa}, Fig.~\ref{fig:manyfits}a, and Fig.~\ref{fig:manyfits}b, the percolation threshold is a fitting parameter $\phi_c$, whereas in Fig.~\ref{fig:manyfits}c and Fig. \ref{fig:manyfits}d, the percolation threshold is held fixed and estimated (for N=50) from the equation $\phi_c(\delta) = 0.035 + 3.67\cdot\delta^{1.1}$, where $\delta=0.04$ and the numerical parameters are taken from the fit in Fig.~\ref{fig:perctrans}d. Fits are done for two different error functions i)
$\rm Error\propto\sum_i\left(\log\kappa_{i}^{num}-\log\kappa_{i}^{fit}\right)^2$ in Fig. \ref{fig:kappa} and Fig.~\ref{fig:manyfits}a and Fig.~\ref{fig:manyfits}c, and ii) $\rm Error\propto\sum_i\left(\kappa_{i}^{num}-\kappa_{i}^{fit}\right)^2$ in Fig. \ref{fig:manyfits}b and Fig.~\ref{fig:manyfits}d. $\rm\kappa_i^{num}$ is a permeability value obtained from LB simulations for the $i^{th}$ packing, whereas $\rm\kappa_i^{fit}$ is a value for a given set of parameters. The results of these fits are in Table~\ref{tab:fits}.

\begin{table*}[]
\centering
\begin{tabular}{|c|cccc|}
\hline\hline
Figure & \multicolumn{1}{c}{Formula} & \multicolumn{1}{c}{Fitting Parameters} & \multicolumn{1}{c}{Error Function} & Parameter Values \\ \hline\hline
Fig. \ref{fig:perctrans}d & $\rm\phi_c(\delta,N) - \phi^*_c(N) = C_N\cdot\delta^{\beta_N}$ & $C_N$,$\phi_c^*(N)$, $\beta_N$ & $\rm\sum_i\left(\phi_{c,i}^{num} - \phi_{c,i}^{fit}\right)^2$ & \makecell{ $\rm C_{16}=3.70\pm0.9$ \\ $\rm C_{32}=3.77\pm0.7$ \\ $\rm C_{50}=3.67\pm1.0$\\ $\phi_c^*(16)=0.064\pm 0.010$\\ $\phi_c^*(32)=0.053\pm 0.020$\\ $\phi_c^*(50)=0.035\pm 0.014$\\ $\beta_{16}=1.2\pm0.2$\\ $\beta_{32}=1.0\pm0.1$\\ $\beta_{50}=1.1\pm0.2$}\\ \hline
{Fig. \ref{fig:TauR123}b} & $\rm\tau_H=C_{\tau}\left(\phi-\phi_c\right)^{-0.38}$ & $C_{\tau}$,$\phi_c$ & $\rm\sum_i\left(\tau_i^{num}-\tau_i^{fit}\right)^2$ & $\phi_c=0.124\pm 0.004$  \\ \hline 
\multirow{3}
 {*}{Fig. \ref{fig:kappa}}  & $\kappa=C_{\kappa}(\phi-\phi_c)^{\gamma+0.76}\phi^2(1-\phi)^{-2}$ & $\rm C_{\kappa}$,$\rm \phi_c$,$\gamma$ &$\rm\sum_i\left(\log\kappa_i^{num}-\log\kappa_i^{fit}\right)^2$ & \makecell{$\phi_c=0.146\pm0.003$\\ $\gamma=0.83\pm 0.17$} \\\cline{5-5}
 & $\kappa=C_{\kappa}(\phi-\phi_c)^{2.52}\phi^2(1-\phi)^{-2}$ & $C_{\kappa}$,$\phi_c$ & $\rm\sum_i\left(\log\kappa_i^{num}-\log\kappa_i^{fit}\right)^2$ & $\phi_c=0.101\pm0.038$ \\\cline{5-5}
 & $\kappa=C_{\kappa}(\phi-\phi_c)^{\bar{e}}$ & $\rm C_{\kappa}$,$\rm \phi_c$,$\bar{e}$ &$\rm\sum_i\left(\log\kappa_i^{num}-\log\kappa_i^{fit}\right)^2$ & \makecell{$\phi_c=0.115\pm0.053$\\ $\bar{e}=3.61\pm 0.30$} \\
 \hline
 \multirow{3}
 {*}{Fig. \ref{fig:manyfits}a}  & $\kappa=C_{\kappa}(\phi-\phi_c)^{\gamma+0.76}\phi^2(1-\phi)^{-2}$ & $\rm C_{\kappa}$,$\rm \phi_c$,$\gamma$ &$\rm\sum_i\left(\log\kappa_i^{num}-\log\kappa_i^{fit}\right)^2$ & \makecell{$\phi_c=0.146\pm0.003$\\ $\gamma=0.83\pm 0.17$} \\\cline{5-5}
 & $\kappa=C_{\kappa}(\phi-\phi_c)^{2.52}\phi^2(1-\phi)^{-2}$ & $C_{\kappa}$,$\phi_c$ & $\rm\sum_i\left(\log\kappa_i^{num}-\log\kappa_i^{fit}\right)^2$ & $\phi_c=0.101\pm0.038$ \\\cline{5-5}
 & $\kappa=C_{\kappa}(\phi-\phi_c)^{\bar{e}}$ & $\rm C_{\kappa}$,$\rm \phi_c$,$\bar{e}$ &$\rm\sum_i\left(\log\kappa_i^{num}-\log\kappa_i^{fit}\right)^2$ & \makecell{$\phi_c=0.115\pm0.053$\\ $\bar{e}=3.61\pm 0.30$} \\
 \hline
  \multirow{3}
{*} {Fig. \ref{fig:manyfits}b} & $\kappa=C_{\kappa}(\phi-\phi_c)^{\gamma+0.76}\phi^2(1-\phi)^{-2}$ & $\rm C_{\kappa}$,$\rm \phi_c$,$\gamma$ &$\rm\sum_i\left(\kappa_i^{num}-\kappa_i^{fit}\right)^2$ & \makecell{$\phi_c=0.150\pm0.002$\\ $\gamma=0.86\pm 0.14$} \\\cline{5-5}
 & $\kappa=C_{\kappa}(\phi-\phi_c)^{2.52}\phi^2(1-\phi)^{-2}$ & $C_{\kappa}$,$\phi_c$ & $\rm\sum_i\left(\kappa_i^{num}-\kappa_i^{fit}\right)^2$ & $\phi_c=0.060\pm0.066$ \\\cline{5-5}
 & $\kappa=C_{\kappa}(\phi-\phi_c)^{\bar{e}}$ & $\rm C_{\kappa}$,$\rm \phi_c$,$\bar{e}$ & $\rm\sum_i\left(\kappa_i^{num}-\kappa_i^{fit}\right)^2$ & \makecell{$\phi_c=0.120\pm0.001$\\ $\bar{e}=3.88\pm 0.18$} \\
 \hline
\multirow{3}
{*}{Fig. \ref{fig:manyfits}c} & $\kappa=C_{\kappa}(\phi-\phi_c)^{\gamma+0.76}\phi^2(1-\phi)^{-2}$ & $\rm C_{\kappa}$,$\gamma$ &$\rm\sum_i\left(\log\kappa_i^{num}-\log\kappa_i^{fit}\right)^2$ & \makecell{$\phi_c(\rm fixed)=0.141$\\ $\gamma=0.91\pm 0.13$} \\\cline{5-5} 
 & $\kappa=C_{\kappa}(\phi-\phi_c)^{2.52}\phi^2(1-\phi)^{-2}$ & $C_{\kappa}$ & $\rm\sum_i\left(\log\kappa_i^{num}-\log\kappa_i^{fit}\right)^2$ & $\phi_c(\rm fixed)=0.141$ \\\cline{5-5}
  & $\kappa=C_{\kappa}(\phi-\phi_c)^{\bar{e}}$ & $\rm C_{\kappa}$,$\bar{e}$ &$\rm\sum_i\left(\log\kappa_i^{num}-\log\kappa_i^{fit}\right)^2$ & \makecell{$\phi_c(\rm fixed)=0.141$\\ $\bar{e}=2.72\pm 0.14$} \\
 \hline
\multirow{3}
 {*}{Fig. \ref{fig:manyfits}d} & $\kappa=C_{\kappa}(\phi-\phi_c)^{\gamma+0.76}\phi^2(1-\phi)^{-2}$ & $\rm C_{\kappa}$,$\gamma$ &$\rm\sum_i\left(\kappa_i^{num}-\kappa_i^{fit}\right)^2$ & \makecell{$\phi_c(\rm fixed)=0.141$\\ $\gamma=0.97\pm 0.16$} \\\cline{5-5} 
 & $\kappa=C_{\kappa}(\phi-\phi_c)^{2.52}\phi^2(1-\phi)^{-2}$ & $C_{\kappa}$ & $\rm\sum_i\left(\kappa_i^{num}-\kappa_i^{fit}\right)^2$ & $\phi_c(\rm fixed)=0.141$ \\\cline{5-5} 
& $\kappa=C_{\kappa}(\phi-\phi_c)^{\bar{e}}$ & $\rm C_{\kappa}$,$\bar{e}$ &$\rm\sum_i\left(\kappa_i^{num}-\kappa_i^{fit}\right)^2$ & \makecell{$\phi_c(\rm fixed)=0.141$\\ $\bar{e}=3.50\pm 0.17$} \\  \hline\hline
\end{tabular}
\caption{Fitting parameters for a percolation threshold in the continuum limit $\phi_c^*$, tortuosity $\rm\tau_H$, and permeability $\kappa$ that are investigated in this paper. In Fig.~\ref{fig:manyfits}c and Fig.~\ref{fig:manyfits}d, $\phi_c$ is fixed, so no standard deviations are given.}
\label{tab:fits}
\end{table*}

\begin{figure}[h]
\includegraphics[width=\columnwidth,keepaspectratio]
{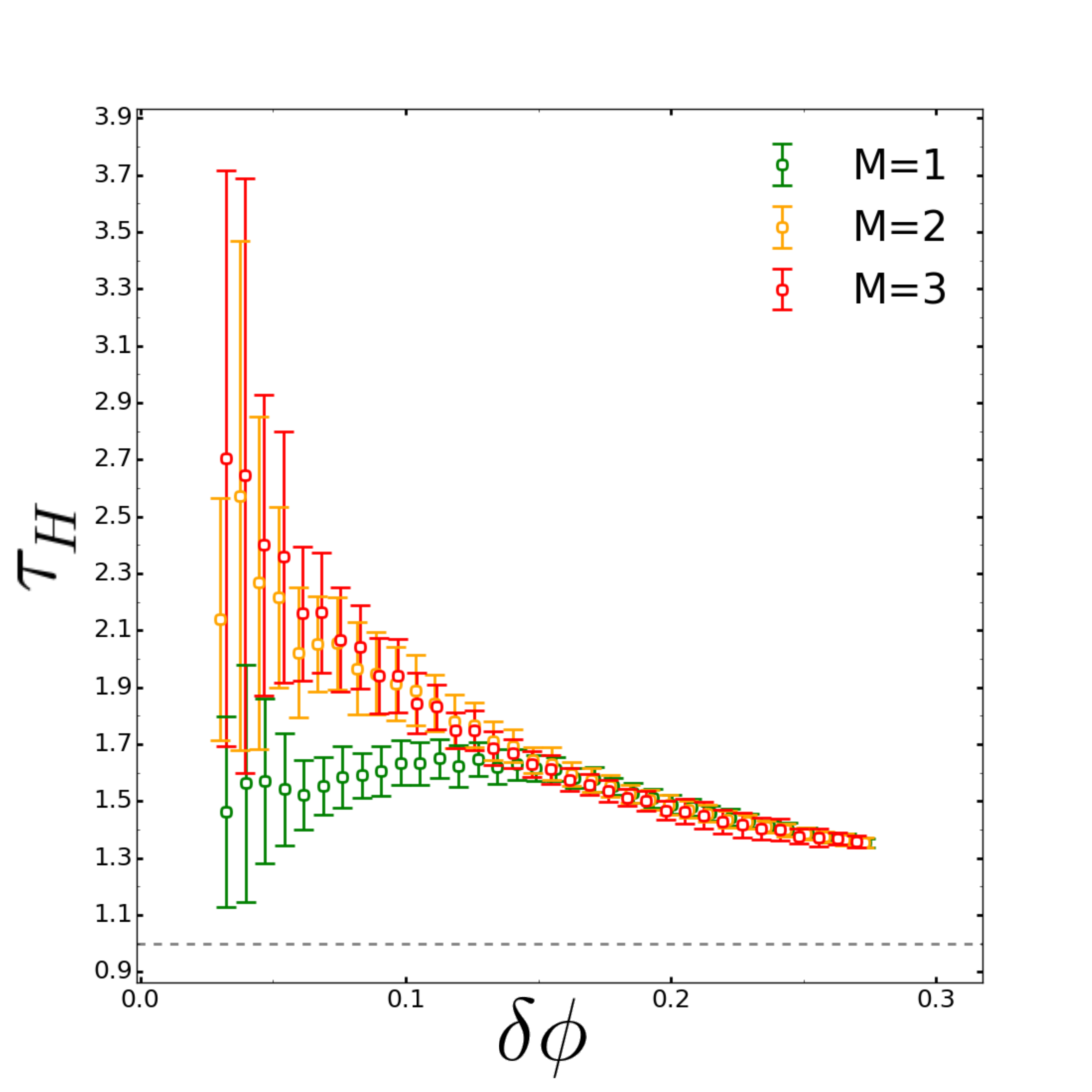}
\caption{The same numerical data as in Fig. \ref{fig:TauR123} \textemdash~with error bars.}
\label{fig:TauR123_error_bars}
\end{figure}

\begin{figure*}[ht]
\includegraphics[width=\textwidth,keepaspectratio]
{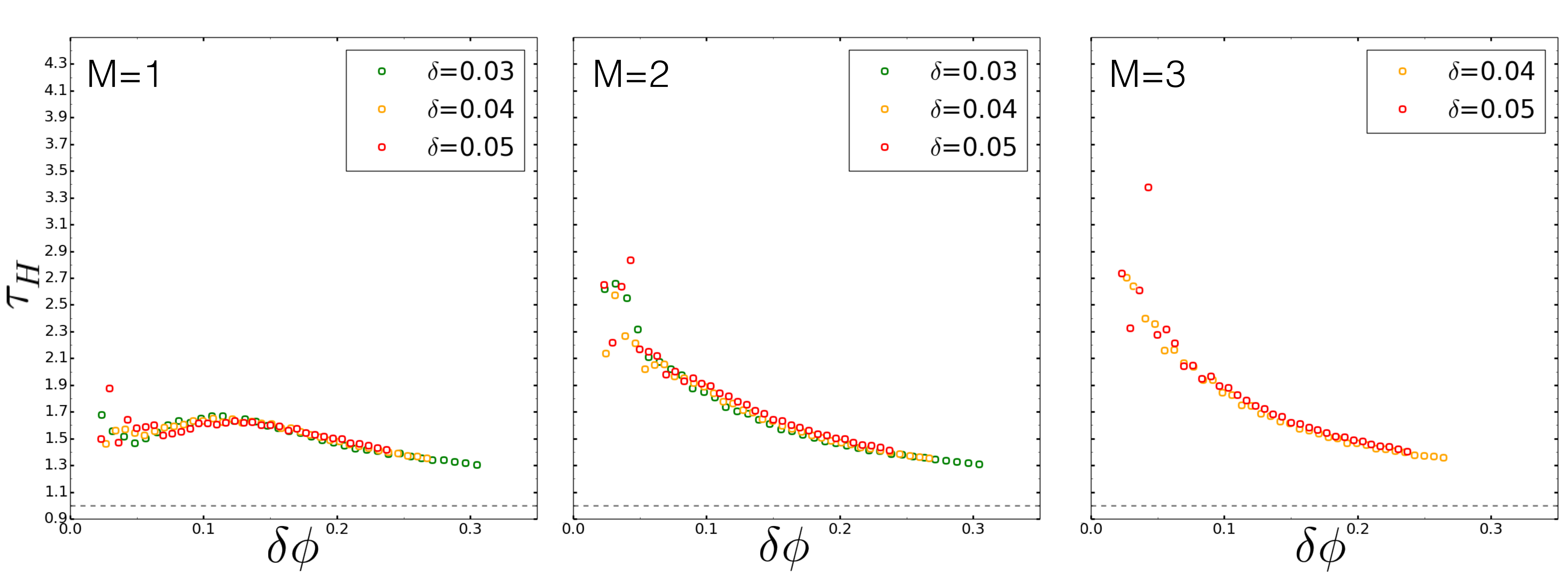}
\caption{\textbf{Tortuosity calculated for three different $\delta$ and different lattice refinement levels}: LB simulations are performed on percolating clusters that are detected for three different lattice sizes: $\delta=[0.03, 0.04, 0.05]$, and $N=50$. The flow fields are resolved on lattices with a size $\rm\tilde{\delta}=\delta/M$, where $\rm M=1,2,3$ are lattice refinement levels \cite{matyka2008tortuosity}. Tortuosity increases as the refinement level increases, consistent with previous studies \cite{matyka2008tortuosity}. The same behavior is observed for all $\delta$. This suggests that the abrupt increase of $\tau_H$ close to the percolation threshold is caused by the fractal geometry of the percolation cluster rather than by artifacts of the numerical methods. Each data-point is an average from about 100 simulations. $\tau_H$ is given as a function of $\delta\phi$, where lattice size dependent percolation threshold was estimated from the fits in Fig.~\ref{fig:perctrans}d, and Table~\ref{tab:fits}: $\{\phi_c(0.03)=0.113,\phi_c(0.04)=0.141,\phi_c(0.05)=0.171\}$. Error-bars are not shown for better readability.}
\label{fig:threedeltas}
\end{figure*}

\bibliographystyle{apsrev4-1} 
\bibliography{references} 

\end{document}